\documentclass[12pt,thmsa]{article}
\usepackage{sw20lart}

\input{tcilatex}
\begin{document}

\smallskip

\begin{center}
\smallskip {\LARGE Universal construction for the unsorted quantum search
algorithms }
\end{center}

\smallskip

\begin{center}
Xijia Miao

Laboratory of Magnetic Resonance and Atomic and Molecular Physics,

Wuhan Institute of Physics and Mathematics, The Chinese Academy of

Sciences, Wuhan 430071, People$^{\prime }$s Republic of China;

Present correspondence address: Center for Magnetic Resonance Research,
University of Minnesota, 2021 6th ST SE, Minneapolis, MN 55455, USA, E-mail:
miao@cmrr.umn.edu

\ 

\textbf{Abstract}
\end{center}

The multiple-quantum operator algebra formalism has been exploited to
construct generally an unsorted quantum search algorithm. The exponential
propagator and its corresponding effective Hamiltonian are constructed
explicitly that describe in quantum mechanics the time evolution of a
multi-particle two-state quantum system from the initial state to the output
of the unsorted quantum search problem. The exponential propagator usually
may not be compatible with the mathematical structure and principle of the
search problem and hence is not a real quantum search network, but it can be
further decomposed into a product of a series of the oracle unitary
operations such as the selective phase-shift operations and the nonselective
unitary operations which can be expressed further as a sequence of
elementary building blocks, i.e., the one-qubit quantum gates and the
two-qubit diagonal phase gates, resulting in that the decomposed propagator
is compatible with the mathematical structure and principle of the search
problem and hence becomes a real quantum search network. The decomposition
for the propagator can be achieved with the help of the operator algebra
structure and symmetry of the effective Hamiltonian, and the properties of
the multiple-quantum operator algebra subspaces, especially the
characteristic transformation behavior of the multiple-quantum operators
under the z-axis rotations. It has been shown that the computational
complexity of the search algorithm is dependent on that of the numerical
multidimensional integration and hence it is believed that the search
algorithm could solve efficiently the unsorted search problem. An NMR device
is also proposed to solve efficiently the unsorted search problem in
polynomial time. \newline
\ \textbf{\newline
1. Introduction}

Quantum computation is a cross discipline among mathematics, quantum physics
and information science. It obeys both the quantum mechanical laws and the
mathematical principles [1-5]. It has been discovered that quantum computers
can solve efficiently certain problems in polynomial time that can not be
solved efficiently by any classical digital computers [3, 6-11]. The famous
examples include the prime factorization [10] and the quantum simulation [3,
11] whose polynomial-time quantum algorithms have been discovered. However,
in practice these problems including the prime factorization and the quantum
simulation are rather specialized. An important question in quantum
computation therefore arises whether or not quantum computers can solved
efficiently a general NP problem in polynomial time [10, 12, 13]. It is well
known that NP-problems are hard for any classical computation. In the past
several years this question has been discussed extensively but has remained
largely ignored [12, 13]. On the other hand, it has been found that a quite
broad class of problems such as search and optimization problems can be
speeded up quadratically by quantum computation [14]. The unsorted search
problem is really a hard problem in classical computation. Assume that there
is a large unsorted database $T=(0,1,...,N-1)$ $(N=2^{n}),$ in which only
one of these elements satisfies the function $f(s)=1,$ but $f(r)=0$\ for any
other element $r$ $(r\neq s)$. Now one wants to find the target element $s$.
If a classical digital computer is used to search for the target element,
one will need to examine $N$ elements of the database in the worst case and
an average of $N/2$ elements before finding the desired element $s$.
However, Grover [14] has showed recently that if a quantum computer is used
one needs to examine only $\sqrt{N}$ elements around to find the target
element. It has been proven that the Grover algorithm is the optimal quantum
search algorithm so far [15]. Grover and his coworkers has shown further how
his search algorithm can speed up quadratically almost any other quantum
algorithms [16, 17]. However, the Grover algorithm is not an efficient
quantum search algorithm and therefore can not solve efficiently a general
NP-problem in polynomial time [12].

One of the most important characteristic features for quantum computation
different from classical counterpart is that quantum computers can have the
ability of the massive parallel processing in computation [4]. In principle,
a quantum computer can offer the possibility for solving efficiently a
general NP problem in polynomial time by virtue of the massive quantum
parallelism. However, the quantum computational output usually can not be
obtained correctly and directly in polynomial time due to the limits of
quantum mechanical measurement. This is the reason why there exists the
question whether a quantum computer can solve efficiently a general
NP-problem or not. As a consequence of the Grover quantum search algorithm
[14], it has been shown that the correct quantum computational output may be
obtained after$\sqrt{N}$ iterations and therefore, a general NP problem may
be speeded up quadratically by quantum computation [12]. It has been
believed extensively that quantum computers can solve efficiently a general
NP-problem by virtue of the massive quantum parallelism, although so far
only few polynomial-time quantum algorithms have been discovered to solve
efficiently some special hard problems [4, 10, 13].

The multiple-quantum operator algebra formalism has been proposed to
describe quantum computational process [18, 19]. It has been exploited
extensively to design the quantum computational network of a known quantum
algorithm [19], construct elementary building blocks of quantum computation
[19, 20a], and prepare the effective pure states in the NMR quantum
computation [20b]. In this paper the multiple-quantum operator algebra
formalism has been used to construct an unsorted quantum search algorithm
and its quantum computational networks that obey the quantum mechanical laws
and are compatible with the mathematical structure and principles of the
unsorted search problem. This quantum search algorithm is different from the
Grover$^{\prime }$s one in that both two algorithms have different
propagators. Its computational complexity is dependent on that of the
numerical multidimensional integration. Therefore, it is believed that the
quantum search algorithm could solve efficiently the unsorted search problem
on an oracle universal quantum computer. This quantum search algorithm is
constructed with two families of elementary unitary operations, that is, the
nonselective unitary operations and the oracle unitary operations, e.g., the
selective phase-shift operations, all these operations can be further
decomposed into a product of a series of elementary building blocks such as
one-qubit quantum gates and the two-qubit diagonal phase gates [19]. Here
assume that the quantum search algorithm runs on a quantum computer with a
quantum system consisting of $n$ two-state particles such as $n$ nuclear
spins with the angular momentum quantum number I=1/2 and one can manipulate
at will each individual two-state particle of the quantum system by an
external field such as an electromagnetic field. Also assume that any
decoherence effects are ignored in the quantum system.\newline
\newline
\textbf{2. The effective Hamiltonian of a quantum system and the
construction of quantum algorithms}

Benioff was the first time to use quantum mechanics to describe the
reversible computation process on classical Turing machines by understanding
the corresponding relationship between the reversible computation process
and the fact that time evolution of an isolated quantum system is reversible
dynamic [1, 2]. He showed that quantum mechanical models were
computationally as powerful as the classical Turing machines. Feynman was
the first person to conjecture that quantum mechanical models might be more
powerful than any classical computers in simulating quantum processes [3].
His universal quantum simulator could efficiently simulate any quantum
dynamics of a quantum system whose Hamiltonian consists of any local
interactions [11]. Deutsch formalized the concept of the universal quantum
computer and has showed that quantum Turing machines could be more powerful
than the classical counterpart from a computational complexity point of view
[4]. He has also developed the quantum circuit model of quantum computation
[5]. Yao has showed further that the two models of quantum computation,
i.e., the quantum Turing machine and the quantum circuit model, are
polynomially equivalent to each other [21]. Therefore, there are two ways of
thinking about quantum computation [4, 5, 8]. One way is to think of it as
the reversible computation on a quantum Turing machine, and another is that
quantum computation can be thought of as the unitary time evolution of a
quantum system from the input state to the output. The unitary time
evolution can be described by a unitary transformation that acts on the
input quantum state in a quantum system. In quantum mechanics, time
evolution of a quantum system may obey merely quantum mechanical laws, e.g.,
the Schr$\ddot{o}$dinger equation. Time-evolutional propagator of a quantum
system characterizes completely the unitary dynamical behaviors of the
quantum system. However, in quantum computation time evolution of a quantum
system during quantum computing from the input quantum state to the output
is also subjected to the mathematical structure and principle of the problem
solved by a given quantum algorithm running on the quantum system in
addition to these quantum mechanical laws [19]. Therefore, the form of the
propagator is constrained by the mathematical structure and principle of the
quantum algorithm. This propagator may be considered generally as a unitary
transformation that may be taken as an exponential unitary operator $U(t)=$
exp$(-iHt)$, where the operator $H$ is the effective Hamiltonian of the
quantum system subjected to the quantum algorithm. This shows that the form
of the effective Hamiltonian $H$ is also constrained by the mathematical
structure and principle of the quantum algorithm. Then the effective
Hamiltonian $H$ could also really characterize the mathematical structure
and properties of the quantum algorithm such as the quantum computational
complexity. For example, provided that the effective Hamiltonian $H$
consists of any local interactions in the quantum system, there could be a
quantum network that can simulate efficiently time evolution of the quantum
system subjected to the quantum algorithm [3, 11]. Obviously, the quantum
network is always compatible with the quantum mechanical laws. Is the
quantum network also compatible with the mathematical structure and
principle of the problem solved by the quantum algorithm? Evidently, if the
quantum network is designed according to the mathematical structure and
principle of the problem it should be a real quantum computational network
to solve the problem, otherwise it can not be thought of as a real quantum
network to solve the problem. Now, if the problem solved by the quantum
algorithm is an NP-problem this real quantum computational network can solve
efficiently the NP-problem because it can simulate efficiently time
evolution of the quantum system subjected to the quantum algorithm.

A quantum computation is a unitary time-evolutional dynamical process
subjected to a given quantum algorithm from the input quantum states to the
output in a quantum system. Now one is given a mathematical problem that
needs to be solved on a quantum computer. Assume that there are a number of
quantum algorithms to solve the same problem. In practice, one may also
design a number of quantum computational networks for a given quantum
algorithm to solve the same problem. Consider the special case that the
initial state and the output in the quantum system are fixed for a given
mathematical problem to be solved on the quantum computer. For example, for
the unsorted quantum search problem the initial state is usually considered
as the superposition and the output state is the target state in a quantum
system. It is well known that there are a number of unitary time-evolutional
pathways from the fixed input state to the output in a quantum system. Each
such pathway is governed by the quantum mechanical laws and described by a
time-evolutional propagator or its corresponding effective Hamiltonian.
There may be a unitary dynamical process of the quantum system subjected to
the quantum algorithm to solve the problem among all these unitary
time-evolutional pathways, and this process is characterized completely by
the effective Hamiltonian during running the quantum algorithm. This unitary
time-evolutional process obeys not only the quantum mechanical laws but also
the mathematical structure and principle of the problem. If there are a
number of quantum algorithms to solve the same problem with the fixed input
state and the output, there is also a unitary dynamical process
corresponding to each such quantum algorithm that obeys both the quantum
mechanical laws and the mathematical principles of the problem. Now, assume
that the problem is an NP-problem in classical computation, is there an
efficient quantum algorithm to solve the problem among these quantum
algorithms? If the efficient quantum algorithm exists how to find it and
construct its quantum computational network? These problems have not be
solved and reminded largely ignored to date. One the other hand, there
already exists a simple scheme to find a quantum algorithm to solve a
mathematical problem when there is a classical algorithm to solve the same
problem. This simple scheme is that a quantum algorithm can be obtained from
its corresponding classical algorithm. The computational network of the
classical algorithm usually consists of a sequence of irreversible classical
logical gates. It can be translated into a quantum algorithm with at least
the same computational power by simply replacing the irreversible classical
logic gates with the corresponding reversible quantum gates according to the
Bennett$^{\prime }$s suggestion [22]. However, this simple scheme usually
may not be available for finding an efficient quantum algorithm to solve a
general NP-problem. In this paper the multiple-quantum operator algebra
formalism is exploited to design a new quantum algorithm and construct its
quantum computational network for a given mathematical problem such as the
unsorted search problem [18], where it is assumed that the input state and
the output of the quantum system for the problem to be solved are given in
advance. Then the total time-evolutional propagator that transforms
unitarily the input state to the output can be constructed explicitly. In
general, the propagator obeys the quantum mechanical laws but usually is not
compatible with the mathematical structure and principle of the problem.
Then it is not a real quantum algorithm to solve the problem. However, the
propagator may be further decomposed into a sequence of the quantum circuit
units which are compatible with the mathematical structure and principle of
the problem with the help of the properties of the multiple-quantum operator
algebra spaces [18, 19] and the operator algebra structure and symmetry of
the effective Hamiltonian. Finally these quantum circuit units are further
decomposed into a product of a series of elementary building blocks, i.e,
the one-qubit quantum gates and the two-qubit diagonal phase gates [19].
Such constructed quantum network is obviously governed by the quantum
mechanical laws and compatible with the mathematical structure and principle
of the problem to be solved. Therefore, it becomes a real quantum algorithm
to solve the problem. Obviously, once the propagator is decomposed into a
product of a polynomial number of elementary building blocks the constructed
quantum network is really a polynomial-time quantum computational network to
solve efficiently the problem no matter whether the problem is an NP-problem
or a polynomial-time problem in classical computation. As an example, a new
quantum search algorithm to find the marked element in an unsorted database
is explicitly constructed with the help of the multiple-quantum operator
algebra formalism. Its quantum computational network consists of the two
types of elementary quantum circuit units, that is, the oracle quantum
unitary operations, e.g., the selective phase-shift operations, and the
nonselective unitary operations, i.e., the oracle-independent quantum
unitary operations. \newline
\newline
\textbf{3. The nonselective unitary operations}

A nonselective unitary operation is an oracle-independent unitary operation
that acts on every two-state particle of a quantum system symmetrically,
that is, all the two-state particles in the quantum system are
indistinguishable and symmetrical with respect to the unitary operation. The
nonselective unitary operations are independent of the marked state to be
searched for in the quantum system and can be implemented on a quantum
computer without knowing in advance any state of the quantum system. They
are also independent of whether the database under search is sorted or
unsorted. Therefore, this type of unitary operations are compatible with the
mathematical structure and principles of the search problem and can be used
to build quantum computational networks of a quantum search algorithm. The
Walsh-Hadamard transformation $W$ and the diffusion transform $D$ in the
Grover algorithm [14] are typical nonselective unitary operations. Besides
the two unitary operations there are a number of other nonselective unitary
operations. As an example, two types of general nonselective unitary
operations are given below, which may be encountered in present quantum
search algorithm. One type of the nonselective rotation operations that are
applied to all the two-state particles in a quantum system symmetrically are
defined by

$\qquad \qquad R_{p}(\theta ,m)=\exp (-i\theta F_{p}^{m})$ $%
(m=1,2,...;p=x,y,z).\qquad \qquad \quad (1)$ \newline
These nonselective unitary operations are constructed with the $m$th power
of the symmetrical Hermitian operator $F_{p}$ defined by (in a spin-1/2
language)

$\qquad \qquad \quad \quad \quad \qquad \ F_{p}=\stackunder{k=1}{\stackrel{n%
}{\sum }}I_{kp}$ $\qquad \quad \quad \quad \qquad \qquad \quad \qquad \qquad
\qquad (2)$ \newline
where the magnetization operator $I_{kp}=\frac{1}{2}\sigma _{kp}\ (p=x,y,z),$
$\sigma _{k}$ is the Pauli$^{^{\prime }}$s operator of the $k$th two-state
particle of the quantum system. Actually, the Walsh-Hadamard transform $W$
can be decomposed into a product of the nonselective unitary operations $%
R_{p}(\theta ,1)$ [19]:

$\qquad \qquad W=\exp (in\pi /2)\exp (-i\pi F_{x})\exp (-i\frac{\pi }{2}%
F_{y}).\qquad \qquad \qquad \qquad \ (3)$ \newline
Another type of nonselective unitary operations are defined by

$\qquad \qquad T(\theta ,\alpha ,\beta _{x},\beta _{y},\beta _{z})=\exp
(-i\theta \stackrel{n}{\stackunder{k=1}{\bigotimes }}A_{k})\qquad \qquad
\qquad \qquad \qquad \ \ (4)$ \newline
where the Hermitian operator $A_{k}$ of the $k$th two-state particle of the
quantum system may be generally chosen as (in a spin-1/2 language)

$\qquad \qquad A_{k}=\alpha E_{k}+2\beta _{x}I_{kx}+2\beta _{y}I_{ky}+2\beta
_{z}I_{kz}$ $\qquad \qquad \qquad \qquad \quad \ \ (5)$ \newline
and the parameter vector $\{\beta _{p}\}$ is a vector with unit magnitude.
The real parameters $\alpha $ and $\beta _{p}$ $(p=x,y,z)$ are independent
of any index $k$, indicating that the unitary operator $T(\theta ,\alpha
,\beta _{x},\beta _{y},\beta _{z})$ is a nonselective unitary operation.
Actually, the nonselective unitary operation can be converted unitarily into
simple nonselective phase-shift operations. One of the nonselective
phase-shift operations is given by [23, 24]

$\qquad C_{0}(\beta )=Diag[e^{-i\beta },1,....,1]$

$\qquad =\exp [-i\beta (\frac{1}{2}E_{1}+I_{1z})\bigotimes (\frac{1}{2}%
E_{2}+I_{2z})\bigotimes ...\bigotimes (\frac{1}{2}E_{n}+I_{nz})]$. \qquad $%
\quad \ \ (6)$ \newline
When $\alpha =1$ the nonselective unitary operation $T(\theta ,\alpha ,\beta
_{x},\beta _{y},\beta _{z})$ (4) can be transferred into the nonselective
phase-shift operation $C_{0}(\beta ).$ Another is defined by

$\qquad S(\beta )=\exp (-i\beta \times 2I_{1z}\bigotimes 2I_{2z}\bigotimes
...\bigotimes 2I_{nz}).\qquad \qquad \qquad \qquad \quad \ \ (7)$ \newline
When $\alpha =0$ the unitary operation $T(\theta ,\alpha ,\beta _{x},\beta
_{y},\beta _{z})$ (4) can be converted unitarily into the nonselective
phase-shift operation $S(\beta ).$ In particular, the general diffusion
transform can be constructed with the nonselective unitary operations
mentioned above:

$\qquad \qquad \qquad \ D(\theta )=-WC_{0}(\theta )W.$ $\qquad \qquad \qquad
\qquad \qquad \qquad \quad \qquad (8)$ \newline
It can prove easily that $D(\theta )=-E+(1-e^{-i\theta })P,$ where $E$ is
unity operator and the project operator $P_{ij}=1/N,$ for all $i$, $j$.
Clearly, the diffusion transform in the Grover algorithm $D=D(\pi )$ [14]$.$%
\newline
\newline
\textbf{4. The selective unitary transformation and the oracle quantum
unitary operation}

The selective unitary operations are related only to the marked state $%
|s\rangle $ that is to be searched for in the quantum system. A type of
particularly important selective unitary operations in an unsorted quantum
search problem are the selective phase-shift unitary operations. For
example, the selective phase inversion operation $C_{s}$ for the marked
state in the Grover search algorithm [14] is a typical selective unitary
operation. This type of the selective phase-shift operations can be defined
as the diagonal unitary operator $C_{s}(\theta )$ ($s\neq 0,$ $N-1$) that
has diagonal unitary representation in usual quantum computational basis:

$\qquad C_{s}(\theta )=Diag\{1,...,1,e^{-i\theta },1,...,1\}=\exp [-i\theta
E_{ss}]\ \ \qquad \qquad \ \ \qquad (9)$\newline
where $[C_{s}(\theta )]_{ss}=e^{-i\theta }$ only for the diagonal index $s$,
and unit for any other diagonal index $t\neq s$. In particular, the
selective phase inversion operation $C_{s}=C_{s}(\pi )$ [14]. As an
exception, $C_{0}(\theta )$ and $C_{N-1}(\theta )$ are two nonselective
phase-shift operations. The diagonal operator $E_{ss}$ of Eq.(9) can be
expressed as

$\qquad E_{ss}=Diag\{0,...,0,1,0,...,0\}$

$\quad \qquad \ \ =(\frac{1}{2}E_{1}+a_{1}^{s}I_{1z})\bigotimes (\frac{1}{2}%
E_{2}+a_{2}^{s}I_{2z})\bigotimes ...\bigotimes (\frac{1}{2}%
E_{n}+a_{n}^{s}I_{nz})\qquad \qquad (10)$ \newline
where $E_{k}$ is the $2\times 2$-dimensional unity operator and $%
a_{k}^{s}=\pm 1,\ k=1,2,...,n.$ It is easy to see that

$\frac{1}{2}E_{k}+a_{k}^{s}I_{kz}=\left( 
\begin{array}{ll}
1 & 0 \\ 
0 & 0
\end{array}
\right) ,$ if $a_{k}^{s}=+1;$ $\frac{1}{2}E_{k}+a_{k}^{s}I_{kz}=\left( 
\begin{array}{ll}
0 & 0 \\ 
0 & 1
\end{array}
\right) ,$ if $a_{k}^{s}=-1.$\newline
When the selective phase-shift operation $C_{s}(\theta )$ acts on an
arbitrary computational basis $|r\rangle $ a phase shift of exp$(-i\theta )$
is generated if and only if $|r\rangle =|s\rangle ,$

$\qquad \qquad \qquad \qquad C_{s}(\theta )|r\rangle =\exp (-i\theta \delta
_{rs})|r\rangle .\qquad \qquad \qquad \qquad \qquad \ (11)$ \newline
Obviously, given an $n$-dimensional unity-number vector $%
\{a_{1}^{s},a_{2}^{s},...,a_{n}^{s}\}$ of the mared state $|s\rangle $ the
diagonal operator $E_{ss}$ is determined uniquely through Eq.(10) and vice
versa. Therefore, the unity-number vector $%
\{a_{1}^{s},a_{2}^{s},...,a_{n}^{s}\}$ characterizes completely the diagonal
operator $E_{ss}$ and also the selective phase-shift operation $C_{s}(\theta
)$.

In general, a selective phase-shift unitary operation may be defined as a
diagonal unitary exponential operator

$\qquad \qquad \qquad \qquad G_{s}(\theta )=\exp (-i\theta \tilde{H}%
_{s})\qquad $ \newline
where the diagonal Hamiltonian $\tilde{H}_{s}=\tilde{H}%
_{s}(a_{1}^{s},a_{2}^{s},...,a_{n}^{s})$ is dependent only upon the
unity-number vector $\{a_{k}^{s}\}$ of the marked state $|s\rangle .$ The
diagonal Hamiltonian $\tilde{H}_{s}$ belongs the \textit{lo}ngitudinal 
\textit{m}agnetization and \textit{s}pin \textit{o}rder ($LOMSO$) operator
subspace of the Liouville operator space of the two-state quantum system and
can be generally expressed as a sum of the base operators of the subspace
[18]

$\tilde{H}_{s}=\Omega _{0}^{^{\prime }}+\stackrel{n}{\stackunder{k=1}{\sum }}%
\Omega _{k}^{^{\prime }}I_{kz}+\stackrel{n}{\stackunder{k<l}{\sum }}%
J_{kl}^{^{\prime }}2I_{kz}I_{lz}+\stackrel{n}{\stackunder{k<l<m}{\sum }}%
J_{klm}^{^{\prime }}4I_{kz}I_{lz}I_{mz}+...$ \qquad $\ \ (12)$\newline
where all the coefficients $\{\Omega _{k}^{^{\prime }},J_{kl}^{^{\prime
}},J_{klm}^{^{\prime }},...\}$ are dependent only on the unity-number vector 
$\{a_{1}^{s},a_{2}^{s},...,a_{n}^{s}\}.$

A selective unitary operation is a black-box operation in a quantum search
problem because this operation is dependent on the marked state $|s\rangle $%
, whereas the marked state needs to be searched for in the quantum system.
Therefore, the selective unitary operations could be implemented only on an
oracle quantum computer. A quantum oracle may be defined as a device that,
when called, applies a fixed unitary transformation $U_{o}$\ to the current
quantum state $|r\rangle $ of the quantum system, replacing it by $%
U_{o}|r\rangle $ [12]. There are some requirements on the unitary
transformation $U_{o}$ in the quantum search problem. The unitary
transformation $U_{o}$ is a selective unitary operation or can be expressed
as a sequence of the selective unitary operations and nonselective unitary
operations. The effective Hamiltonian of a quantum oracle corresponding to
the oracle unitary operation $U_{o}(\theta )=\exp (-i\theta H_{o})$ is
therefore expressed as the form

$\qquad \qquad \qquad \qquad
H_{o}=H_{o}(a_{1}^{s},a_{2}^{s},...,a_{n}^{s}).\qquad \qquad \qquad \qquad
\qquad \qquad (13)$ \newline
Another natural restriction to impose upon $U_{o}$ may be that $U_{o}$ is
periodic, that is, $U_{o}^{r}=E$, $r$ is a known integer, so that the effect
of an oracle call can be undone by further $r-1$ calls, two oracle calls
undone by further $r-2$ calls, and so forth, on the same oracle. The key
property of a quantum oracle is its block-box nature [7]. Therefore, an
oracle unitary operation could be implemented only on an oracle quantum
computer. There are a variety of oracle unitary operations in a quantum
search problem. For example, the selective phase inversion operation $%
C_{s}(\pi )$ is chosen as an oracle unitary operation in the Grover
algorithm whose effective Hamiltonian $E_{ss}$ is given in Eq.(10) and the
phase angle $\theta =\pi $ [14]. As a generalization of the selective phase
inversion operation $C_{s}(\pi )$ one can choose conveniently the selective
phase-shift operation $C_{s}(\theta )$ as the oracle unitary operation. The
oracle unitary operation $C_{s}(\theta )$ can be really implemented directly
on an oracle universal quantum computer and by using this oracle unitary
operation one can construct the unsorted quantum search network, as shown
below. Besides the selective phase-shift operation $C_{s}(\theta )$ the
selective unitary operation $U_{op}(\theta )=\exp (-i\theta H_{op})$ whose
effective Hamiltonian is given by

$\qquad H_{op}=\stackrel{n}{\stackunder{k=1}{\sum }}a_{k}^{s}I_{kp}\equiv
(a_{1}^{s}I_{1p})\bigotimes E_{2}\bigotimes ...\bigotimes E_{n}$

$\qquad \qquad +E_{1}\bigotimes (a_{2}^{s}I_{2p})\bigotimes E_{3}\bigotimes
...\bigotimes E_{n}$

$\qquad +...+E_{1}\bigotimes ...\bigotimes E_{n-1}\bigotimes
(a_{n}^{s}I_{np})\ \qquad (p=x,y,z)\qquad \qquad \qquad \ \ (14)$ \newline
also may be a suitable oracle unitary operation used to construct the
quantum search network because it may be implemented directly on an NMR
quantum computer, as shown in Appendix A and C. The effective Hamiltonian $%
H_{op}$ is traceless and the selective unitary operation $U_{op}(\theta )$
of the Hamiltonian $H_{op}$ can be expressed as

$\qquad \qquad \qquad \qquad U_{op}(\theta )=\stackrel{n}{\stackunder{k=1}{%
\prod }}\exp [-i\theta a_{k}^{s}I_{kp}].$\qquad $\qquad \qquad \qquad \quad
\ \ \ \ (15)$\newline
Note that the three oracle unitary operations with different $p=x,y,z$ are
equivalent to each other by a simple nonselective unitary transformation,
for example, $U_{oy}(\theta )=\exp (i\frac{\pi }{2}F_{x})U_{oz}(\theta )\exp
(-i\frac{\pi }{2}F_{x})$ and $U_{ox}(\theta )=\exp (-i\frac{\pi }{2}F_{y})$ $%
\times U_{oz}(\theta )\exp (i\frac{\pi }{2}F_{y}).$ Obviously, $%
[U_{op}(\theta )]^{r}=E$ when $r\theta =4\pi .$ The oracle unitary operation
is really single-qubit $p$-axis pulse applied to all the two-state particles
in the quantum system and the phase of the pulse applied to the $k$th
two-state particle is taken as $(-\theta a_{k}^{s})$. It is well known that
single-qubit quantum operations are always implemented easily in quantum
computation. Therefore, this oracle unitary operation is very simple. By
exploiting the oracle unitary operation $U_{op}(\theta )$ one can decomposed
the selective phase-shift operation $C_{s}(\theta )$ into a product of a
polynomial number of the oracle unitary operations $U_{op}(\theta )$ and the
nonselective unitary operations

$\qquad C_{s}(\theta )=\exp (-i\frac{\pi }{2}F_{y})U_{oy}(-\frac{\pi }{2}%
)C_{0}(\theta )U_{oy}(\frac{\pi }{2})\exp (i\frac{\pi }{2}F_{y}).$\qquad $%
\qquad \quad \ \ (16)$\newline
Obviously, if the oracle unitary operation $U_{op}(\theta )$ could be
implemented efficiently on an oracle quantum computer the selective
phase-shift operation $C_{s}(\theta )$ could be performed in a polynomial
time on the same oracle quantum computer. There is a question whether the
oracle unitary operation $U_{op}(\theta )$ can be expressed as a sequence of
the selective phase-shift operations $C_{s}(\theta )$ and the nonselective
unitary operations. This question will be discussed in the following section
7.

How to implement the selective phase-shift unitary operation $C_{s}(\theta
)? $ Assume that the quantum system is at an arbitrary state including the
marked state $|s\rangle $:

$\qquad |\Psi \rangle =|r,S\rangle =\ \stackrel{N-1}{\stackunder{x=0}{\sum }}%
a_{x}|x\rangle [\frac{1}{\sqrt{2}}(|0\rangle -|1\rangle )]$ \qquad \newline
where the ancillary qubit $S$ is at the superposition $\frac{1}{\sqrt{2}}%
(|0\rangle -|1\rangle ).$ The evaluation of the function $f(x)$ then can be
achieved by performing the oracle unitary operation $U_{f}$ on the state $%
|\Psi \rangle $ [12]

$U_{f}:\ \ |\Psi \rangle \rightarrow \ \stackrel{N-1}{\stackunder{x=0}{\sum }%
}a_{x}|x\rangle [\frac{1}{\sqrt{2}}(|0\bigoplus f(x)\rangle -|1\bigoplus
f(x)\rangle )]$

$\qquad \qquad =\ \stackrel{N-1}{\stackunder{x=0}{\sum }}(-1)^{f(x)}a_{x}|x%
\rangle [\frac{1}{\sqrt{2}}(|0\rangle -|1\rangle )]$

$\qquad \qquad =\ \stackrel{N-1}{\stackunder{x=0,x\neq s}{\sum }}%
a_{x}|x\rangle [\frac{1}{\sqrt{2}}(|0\rangle -|1\rangle )]-a_{s}|s\rangle [%
\frac{1}{\sqrt{2}}(|0\rangle -|1\rangle )]$ \newline
where the function $f(s)=1$ and $f(r)=0$, $r\neq s$ . Obviously, only the
target state $a_{s}|s\rangle [\frac{1}{\sqrt{2}}(|0\rangle -|1\rangle )]$ is
inverted and any other state keeps unchanged when performing once evaluation
of the function $f(x).$ Therefore, performing once evaluation of the
function $f(x)$ is actually equivalent to applying the selective
phase-inversion operation $C_{s}(\pi )$ to the quantum system. The general
selective phase-shift operation $C_{s}(\theta )$ can be achieved by the
following oracle unitary operations

$U_{f}:\ \ |\Psi \rangle =\ \stackrel{N-1}{\stackunder{x=0}{\sum }}%
a_{x}|x\rangle |0\rangle |1\rangle \rightarrow \ \stackrel{N-1}{\stackunder{%
x=0}{\sum }}a_{x}|x\rangle |0\bigoplus f(x)\rangle |1\rangle $

$V(\theta ):\qquad \qquad \rightarrow \ \stackrel{N-1}{\stackunder{x=0}{\sum 
}}\exp [-i\theta \delta (f(x),1)]a_{x}|x\rangle |0\bigoplus f(x)\rangle
|1\rangle $

$U_{f}^{-1}:\qquad \qquad \rightarrow \ \stackrel{N-1}{\stackunder{x=0,x\neq
s}{\sum }}a_{x}|x\rangle |0\rangle |1\rangle +\exp (-i\theta )a_{s}|s\rangle
|0\rangle |1\rangle $ \newline
where the function $\delta (f(x),1)=1$ if $f(x)=1$; otherwise $\delta
(f(x),1)=0,$ and two ancillary qubits are used in the implementation of the
selective phase-shift operation $C_{s}(\theta )$. Therefore, the oracle
unitary operation $C_{s}(\theta )$ can be expressed as

$\qquad \qquad \qquad \qquad C_{s}(\theta )=U_{f}^{-1}V(\theta )U_{f}\qquad
\qquad \qquad \qquad \qquad \quad \ (17)$ \newline
where $V(\theta )$ is a conditional phase-shift operation applying only to
the two ancillary qubits [10, 25]. Note that $U_{f}^{2}=1$ the selective
phase-shift operation can be also expressed as $C_{s}(\theta )=U_{f}V(\theta
)U_{f}$. \newline
\newline
\textbf{5. Construction of quantum search networks}

Assume that each usual quantum computational base $|r\rangle $ of the
quantum system corresponds one-to-one to an element of the search database,
and in particular, the target element is represented by the marked base $%
|s\rangle .$ A usual quantum computational basis can be taken as a Kronecker
product of the common eigenbase of the single-particle spin angular momentum
operators $I_{k}^{2}$ and $I_{kz}$ in a two-state multiparticle quantum
system, for example,

$\qquad \qquad |s\rangle =|\alpha \beta ...\alpha \rangle =|\alpha \rangle
\bigotimes |\beta \rangle \bigotimes ...\bigotimes |\alpha \rangle .$\newline
In the binary representation the eigenbase $|\alpha \rangle $, $|\beta
\rangle $ are denoted briefly as $|0\rangle ,$ $|1\rangle $, respectively.
In the spinor or vector representation the eigenbase can be expressed as $%
|\alpha \rangle =\binom{1}{0}$ and $|\beta \rangle =\binom{0}{1}.$ In the
unity-number representation $\{a_{1}^{s},a_{2}^{s},...,a_{n}^{s}\}$ an
arbitrary usual computational basis $|s\rangle $ can be expressed explicitly
as

$|s\rangle =(\frac{1}{2}T_{1}+a_{1}^{s}S_{1})\bigotimes (\frac{1}{2}%
T_{2}+a_{2}^{s}S_{2})\bigotimes ...\bigotimes (\frac{1}{2}%
T_{n}+a_{n}^{s}S_{n})\qquad \qquad \qquad \ (18)$ \newline
where $T_{k}=|\alpha \rangle _{k}+|\beta \rangle _{k}$ and $2S_{k}=|\alpha
\rangle _{k}-|\beta \rangle _{k}.$ This shows that the unity-number vector $%
\{a_{1}^{s},a_{2}^{s},...,a_{n}^{s}\}$ really characterizes completely the
usual quantum computational basis $|s\rangle $ and vice versa. Obviously, $%
\frac{1}{2}T_{k}+a_{k}^{s}S_{k}=|\alpha \rangle $ if $a_{k}^{s}=1;$ $\frac{1%
}{2}T_{k}+a_{k}^{s}S_{k}=|\beta \rangle $ if $a_{k}^{s}=-1.$

A general quantum search algorithm to find the marked state in the quantum
system may be thought of as a unitary time-evolutional process of the
quantum system from the initial superposition $|\Psi \rangle $ into the
marked state $|s\rangle $

$\qquad \qquad U_{S}:\quad |\Psi \rangle =\dfrac{1}{\sqrt{N}}\stackrel{N-1}{%
\stackunder{k=0}{\sum }}|k\rangle \rightarrow |s\rangle \qquad \qquad \qquad
\qquad \qquad \qquad (19)$ \newline
where the unitary operator $U_{S}$ transforms the initial state to the
output of the quantum system and may be taken as the quantum computational
network of a quantum search algorithm. There are a number of unitary
transformations $U_{S}$ and different unitary transformations $U_{S}$ may
correspond to different quantum search algorithms. In the Grover search
algorithm [14] the unitary transformation $U_{S}$ is taken as a sequence of
the $O(\sqrt{N})$ number of the simple unitary transformations: $-WC_{0}(\pi
)WC_{s}(\pi ),$ where the oracle unitary operation is the selective
phase-inversion operation $C_{s}(\pi ).$ Therefore, the Grover search
algorithm is a quadratically speed-up unsorted quantum search algorithm.
Now, a simple unitary transformation $U_{S}$ is constructed explicitly that
is different from the Grover$^{\prime }$s one and its corresponding
effective Hamiltonian is local. First of all, there is a unitary
transformation that transforms unitarily the computational basis $|r\rangle $
to the marked state $|s\rangle $

$\qquad \qquad \qquad \qquad \qquad U_{rs}|r\rangle =|s\rangle .\qquad
\qquad \qquad \qquad \qquad \qquad \qquad \ \ (20)$ \newline
This unitary transformation $U_{rs}$ may be simply constructed by

$\qquad \qquad \qquad U_{rs}=E-E_{rr}-E_{ss}+2I_{x}^{rs}\qquad \qquad \qquad
\qquad \qquad \qquad (21)$\newline
where $E$ is the unity operator, the operator $E_{rs}$ is defined by

$\qquad \qquad \qquad \qquad (E_{rs})_{ij}=\delta _{ri}\delta _{sj},\qquad
\qquad \qquad \qquad \qquad \qquad \qquad \ \ \ (22)$ \newline
and the single-transition operators are defined by [23]

$\qquad I_{x}^{rs}=\frac{1}{2}(E_{rs}+E_{sr})$, $I_{y}^{rs}=\frac{1}{2i}%
(E_{rs}-E_{sr}),$ $I_{z}^{rs}=\frac{1}{2}(E_{rr}-E_{ss}).$\newline
It can prove that the unitary operator $U_{rs}$ can be further expressed as
the exponential form [18]

$\qquad \qquad \qquad U_{rs}=C_{s}(\pi )\exp (i\pi I_{y}^{rs}).\qquad \qquad
\qquad \qquad \qquad \qquad \ \quad \ \ (23)$ \newline
Therefore, one of the unitary transformations $U_{S}$ of Eq.(19) may be
expressed as

$\qquad \qquad U_{S}=U_{0s}W=C_{s}(\pi )\exp (i\pi I_{y}^{0s})W\qquad \qquad
\qquad \qquad \qquad \ \ \quad (24)$ \newline
where the following transformation involved in the Walsh-Hadamard operation $%
W$ has be introduced [26]

$\qquad \qquad \qquad \qquad W\dfrac{1}{\sqrt{N}}\stackrel{N-1}{\stackunder{%
k=0}{\sum }}|k\rangle =|0\rangle .\qquad \qquad \qquad \qquad \qquad \qquad
\ \ (25)$ \newline

The exponential unitary operator $\exp (i\pi I_{y}^{rs})$ could be prepared
and performed directly on a quantum computer only when the pair quantum
computational base $|r\rangle $ and $|s\rangle $ are known in advance.
However, the basis state $|s\rangle $ is the marked state that needs to be
searched for in a quantum system. Obviously, this operation is not
compatible with the mathematical structure and principle of the search
problem. Then the unitary transformation $U_{S}$ of Eq.(24) is not a real
quantum search network. In order to construct a real quantum search
algorithm one needs to find another unitary transformation that is entirely
equivalent to the unitary operation $\exp (i\pi I_{y}^{0s}).$ This unitary
transformation is required to be compatible with the mathematical structure
and principle of the search problem. Obviously, if the unitary operation $%
\exp (i\pi I_{y}^{0s})$ can be expressed as a sequence of the nonselective
unitary operations and the selective unitary transformations, i.e., the
oracle quantum unitary operations, the unitary transformation $U_{S}$ of
Eq.(24) will be compatible with the mathematical structure and principle of
the quantum search problem and hence becomes a real quantum search network.
Obviously, this quantum search algorithm is different from the Grover$%
^{\prime }$s one that has a different propagator transforming the initial
state to the final state of the search problem [17]. In the following it is
shown that it is possible to express the unitary operator $\exp (i\pi
I_{y}^{0s})$ as a sequence of the nonselective unitary transformations and
the selective unitary operations, and each of these unitary operations can
be further expressed as a sequence of one-qubit quantum gates and the
two-qubit diagonal phase gates.

Now the single-transition multiple-quantum operator $I_{y}^{rs}$ is
expressed as

$\qquad \qquad \qquad \qquad I_{y}^{rs}=\frac{1}{2i}%
(E_{rr}QE_{ss}-E_{ss}QE_{rr})\qquad \qquad \qquad \qquad \quad \ \ (26)$ 
\newline
where the operator $Q$ is defined by the matrix elements $Q_{ij}=1$ for all
indexes $i,$ $j$, and can be written in the unity-number representation as
the form [20b]

$\qquad Q=2^{n}(\frac{1}{2}E_{1}+I_{1x})\bigotimes $ $(\frac{1}{2}%
E_{2}+I_{2x})\bigotimes ...\bigotimes (\frac{1}{2}E_{n}+I_{nx}).\qquad
\qquad \ \ (27)$\newline
By exploiting the operators $E_{rr},$ $E_{ss},$ and $Q$ in the unity-number
representation one obtains that

$\qquad \qquad \qquad \qquad E_{rr}QE_{ss}=\stackrel{n}{\stackunder{k=1}{%
\bigotimes }}f_{k}(a_{k}^{s},a_{k}^{r}),\qquad \qquad \qquad \qquad \qquad \
\ (28a)$

$\qquad \qquad \qquad \qquad E_{ss}QE_{rr}=\stackrel{n}{\stackunder{k=1}{%
\bigotimes }}g_{k}(a_{k}^{s},a_{k}^{r})\qquad \qquad \qquad \qquad \qquad \
\ \ (28b)$ \newline
where

$f_{k}(a_{k}^{s},a_{k}^{r})=[\frac{1}{4}(1+a_{k}^{r}a_{k}^{s})E_{k}+\frac{1}{%
2}(a_{k}^{r}+a_{k}^{s})I_{kz}$

$\qquad \qquad \qquad +\frac{1}{2}(1-a_{k}^{r}a_{k}^{s})I_{kx}+\frac{1}{2}%
i(a_{k}^{r}-a_{k}^{s})I_{ky}],$

$g_{k}(a_{k}^{s},a_{k}^{r})=[\frac{1}{4}(1+a_{k}^{r}a_{k}^{s})E_{k}+\frac{1}{%
2}(a_{k}^{r}+a_{k}^{s})I_{kz}$

$\qquad \qquad \qquad +\frac{1}{2}(1-a_{k}^{r}a_{k}^{s})I_{kx}-\frac{1}{2}%
i(a_{k}^{r}-a_{k}^{s})I_{ky}].$\newline
In particular, $a_{k}^{r}=+1,\ k=1,2,...,n$ for the case of $|r\rangle
=|0\rangle $ and

$f_{k}(a_{k}^{s},a_{k}^{0})=\{\QDATOP{\frac{1}{2}E_{k}+I_{kz}\text{, if }%
a_{k}^{s}=1}{I_{k}^{+},\text{ \ \ \quad \quad if }a_{k}^{s}=-1},$ \newline

$g_{k}(a_{k}^{s},a_{k}^{0})=\{\QDATOP{\frac{1}{2}E_{k}+I_{kz}\text{, if }%
a_{k}^{s}=1}{I_{k}^{-},\text{ \ \ \quad \quad if }a_{k}^{s}=-1}$ \newline
where $I_{k}^{\pm }=I_{kx}\pm iI_{ky}$. Then, by inserting Eq.(28) into
Eq.(26) the multiple-quantum operator $I_{y}^{0s}$ can be rewritten
generally as

$I_{y}^{0s}=\frac{1}{2i}(I_{m_{1}}^{+}I_{m_{2}}^{+}...I_{m_{k}}^{+}$

$\qquad -$ $I_{m_{1}}^{-}I_{m_{2}}^{-}...I_{m_{k}}^{-})(\frac{1}{2}%
E_{m_{k+1}}+I_{m_{k+1}z})...(\frac{1}{2}E_{m_{n}}+I_{m_{n}z})\ \qquad \qquad
\qquad (29)$\newline
where the Kronecker product symbol $\bigotimes $ and index order are omitted
without confusion and here assume that $a_{m_{i}}^{s}=-1$, $i=1,2,...,k;$ $%
a_{m_{i}}^{s}=1,$ $i=k+1,...,n;$ $m_{i}=1,2,...,n.$ For example,

$\frac{1}{2i}(I_{1}^{+}I_{3}^{+}-I_{1}^{-}I_{3}^{-})(\frac{1}{2}%
E_{2}+I_{2z})...=\frac{1}{2i}I_{1}^{+}\bigotimes (\frac{1}{2}%
E_{2}+I_{2z})\bigotimes I_{3}^{+}\bigotimes ...$

$\qquad \qquad \qquad \qquad \qquad \qquad \qquad -\frac{1}{2i}%
I_{1}^{-}\bigotimes (\frac{1}{2}E_{2}+I_{2z})\bigotimes I_{3}^{-}\bigotimes
....$\newline
Equation (29) shows that the operator $I_{y}^{0s}$ is a k-order
multiple-quantum coherence operator $(0<k\leq n)$ [23]. Now one makes a
unitary transformation on the operator $I_{y}^{0s}$ to convert it into the
diagonal operator [20b],

$\qquad \qquad \qquad \qquad UI_{y}^{0s}U^{+}=\frac{1}{2}E_{00}-\frac{1}{2}%
E_{ss}\qquad \qquad \qquad \qquad \qquad \qquad (30)$\newline
where the unitary transformation $U$ turns out to be taken as the form

$\qquad \qquad U=\exp (-i\frac{\pi }{4}\times
2^{n}I_{m_{1}x}...I_{m_{k}x}I_{m_{k+1}z}...I_{m_{n}z}).$\qquad $\qquad
\qquad \quad \ \ (31)$\newline
This unitary operator can be further expressed as a sequence of nonselective
unitary operations and the selective unitary operation $U_{oy}(\pm \frac{\pi 
}{4}):$\newline
$U=\exp (-i\frac{\pi }{4}F_{y})U_{oy}(-\frac{\pi }{4})\exp (-i\frac{\pi }{4}%
\times 2^{n}I_{1z}I_{2z}...I_{nz})U_{oy}(\frac{\pi }{4})\exp (i\frac{\pi }{4}%
F_{y}).\quad \ \ \ (32)$ \newline
It follows from Eq.(30) that the unitary transformation $\exp (i\pi
I_{y}^{0s})$ can be expressed explicitly as

$\qquad \qquad \qquad \exp (i\pi I_{y}^{0s})=U^{+}C_{0}(-\pi /2)C_{s}(\pi
/2)U$ \qquad $\qquad \qquad \qquad (33)$\newline
where the unitary operation $C_{0}(\pi /2)$ is a nonselective unitary
operation and $C_{s}(\theta )$ is the selective phase-shift operation. The
unitary operator $U$ (and $U^{+}$) consists of the selective unitary
operation $U_{oy}(\pm \frac{\pi }{4})$ and the nonselective unitary
operations $\exp (\pm i\frac{\pi }{4}F_{y})$ and $\exp (\pm i\frac{\pi }{4}%
\times 2^{n}I_{1z}I_{2z}...I_{nz})$, as shown in Eq.(32). Inserting Eq.(33)
into Eq.(24) one writes the unitary transformation $U_{S}$ as the form

$\qquad \qquad U_{S}=C_{s}(\pi )U^{+}C_{0}(-\pi /2)C_{s}(\pi /2)UW.\qquad
\qquad \qquad \qquad \quad \ \ (34)$ \newline
Now, is this unitary operator $U_{S}$ a real quantum search network? Because
the selective phase-shift operation $C_{s}(\theta )$ can be expressed as a
simple sequence of the oracle unitary operation $U_{oy}(\pm \frac{\pi }{4})$
and the nonselective unitary operations $\exp (\pm i\frac{\pi }{2}F_{y})$
and $C_{0}(\theta )$, as shown in Eq.(16), the unitary operator $U_{S}$
consists of the oracle unitary operation $U_{oy}(\pm \frac{\pi }{4})$ in
addition to those nonselective unitary operations. Then the quantum network $%
U_{S}$ of Eq.(34) will be a real quantum search network only when the oracle
unitary operation $U_{oy}(\pm \frac{\pi }{4})$ can be directly implemented
on an oracle quantum computer. I will show in Appendix A and C how an NMR
device can be exploited to implement directly and efficiently the oracle
unitary operation $U_{oy}(\pm \frac{\pi }{4}),$ then the quantum network $%
U_{S}$ of Eq.(34) becomes a real efficient quantum search network on the NMR
device. On the other hand, it has been shown in Eq.(17) in section 4 that
the selective phase-shift operation $C_{s}(\theta )$ can be implemented
directly on an oracle universal quantum computer [12, 14]. Then, the quantum
network $U_{S}$ of Eq.(34) will be a real quantum search network when the
oracle unitary operation $U_{oy}(\pm \frac{\pi }{4})$ is expressed
explicitly as a sequence of the selective phase-shift operations $%
C_{s}(\theta )$ and the nonselective unitary operations. Obviously, this
real quantum search network $U_{S}$ (34) is independent of any specific
quantum computer such as an NMR quantum computer and consists of the
selective phase-shift operations $C_{s}(\theta )$ and the nonselective
unitary operations such as the Walsh-Hadamard transform $W$, $\exp (\pm i%
\frac{\pi }{4}\times 2^{n}I_{1z}I_{2z}...I_{nz}),$ etc., which can be
decomposed further into a product of a polynomial number $O(n)$ of one-qubit
quantum gates and the two-qubit diagonal phase gates [19], as can be seen in
next sections. \newline
\newline
\textbf{6. The parallel quantum search networks}

As shown in next sections, there is a complex expression for the oracle
unitary operation $U_{oy}(\theta )$ as the selective phase-shift operations $%
C_{s}(\theta )$ and the nonselective unitary operations. There are four
oracle unitary operations $U_{oy}(\theta )$ in the quantum network $U_{S}$
of Eq.(34). Therefore, the present quantum network is quite complicated and
is not highly efficient. To simplify it one first expands the unitary
operator of Eq.(32) as

$\qquad U=\frac{1}{\sqrt{2}}[E-i\exp (-i\frac{\pi }{2}F_{y})U_{oy}(-\frac{%
\pi }{2})(-1)^{n}2^{n}I_{1z}I_{2z}...I_{nz}]$ \quad $\qquad \qquad (35)$%
\newline
where the following operator identity has been introduced

$\qquad \exp (\pm i\pi F_{p})=(\pm i)^{n}2^{n}I_{1p}I_{2p}...I_{np}\quad
(p=x,y,z).\qquad \qquad \qquad \quad (36)$ \newline
With the help of Eq.(35) the quantum network $U_{S}$ of Eq.(34) can be
expanded as

$U_{S}=\frac{1}{2}\{U_{a}W+U_{a}^{+}W-(-i)^{n+1}C_{s}(\pi )U_{b}C_{0}(-\frac{%
\pi }{2})W$

$\qquad \qquad -(i)^{n+1}C_{0}(-\frac{\pi }{2})U_{b}^{+}W\}\qquad \qquad
\qquad \qquad \qquad \qquad \qquad \qquad \ (37)$ \newline
where

$U_{a}=C_{0}(-\frac{\pi }{2})C_{s}(-\frac{\pi }{2}),$

$U_{b}=\exp (-i\pi F_{z})\exp (i\frac{\pi }{2}F_{y})U_{oy}(\frac{\pi }{2}%
)C_{s}(\frac{\pi }{2}).$ \newline
Equation (37) shows that the quantum network $U_{S}$ is the sum of the four
unitary transformations: $U_{a}W,$ $U_{a}^{+}W,$ $C_{s}(\pi )U_{b}C_{0}(-%
\frac{\pi }{2})W,$ $C_{0}(-\frac{\pi }{2})U_{b}^{+}W,$ indicating that the
quantum network $U_{S}$ could be achieved if one performs in parallelism the
four unitary transformations on the same initial superposition,
respectively, and then sums up coherently their outputs according to
Eq.(37). Obviously, the parallel quantum network of Eq.(37) is more
efficient with respect with the one in Eq.(34) because one needs to perform
in parallelism only once the oracle unitary operation $U_{oy}(\frac{\pi }{2}%
) $.\newline
\newline
\textbf{7. The transformation between oracle unitary operations }

In a quantum search of an unsorted database the selective phase-shift
operations $C_{s}(\theta )$ can be generally implemented directly on an
oracle universal quantum computer, as shown in section 4. Therefore, one
needs to express any other oracle unitary operations such as the oracle
unitary operation $U_{op}(\theta )$ ($p=x,y,z$) in the quantum network $%
U_{S} $ of Eq.(34) as a product of a series of the selective phase-shift
operations $C_{s}(\theta )$ and the nonselective unitary operations.
Obviously, the quantum network $U_{S}$ of Eq.(34) is also real quantum
search algorithm when the oracle unitary operation $U_{op}(\theta )$ is
expressed as a sequence of the selective phase-shift operations $%
C_{s}(\theta )$ and the nonselective unitary operations. How to implement
directly the oracle unitary operation $U_{op}(\theta )$ through an NMR
device in polynomial time is given in Appendix A and C. It is discussed
below how to construct the oracle unitary operation $U_{op}(\theta )$ with
the selective phase-shift operations $C_{s}(\theta )$ and the nonselective
unitary operations.

Now the diagonal operator $E_{ss}$ of Eq.(10)\ is expanded as

$2^{n}E_{ss}=E+\stackrel{n}{\stackunder{k=1}{\sum }}(2a_{k}^{s}I_{kz})+%
\stackrel{n}{\stackunder{1=k<l}{\sum }}(2a_{k}^{s}I_{kz})(2a_{l}^{s}I_{lz})$

$\qquad \qquad +\stackrel{n}{\stackunder{1=k<l<m}{\sum }}%
(2a_{k}^{s}I_{kz})(2a_{l}^{s}I_{lz})(2a_{m}^{s}I_{mz})+....\qquad \qquad
\qquad \qquad \ (38)$\newline
By using the following spin-echo sequence [23] the even-body interactions
such as $(2a_{k}^{s}I_{kz})(2a_{l}^{s}I_{lz}),$ $%
(2a_{k}^{s}I_{kz})(2a_{l}^{s}I_{lz})(2a_{m}^{s}I_{mz})(2a_{n}^{s}I_{nz}),$
etc., on the right-hand of Eq.(38) are cancelled. One therefore has

$U_{ss}\equiv \exp (-i\theta H_{ss})$

$=\exp (-i\theta 2^{n}E_{ss})\exp (-i\pi F_{y})\exp (i\theta
2^{n}E_{ss})\exp (i\pi F_{y})$

$=\exp [-i2\theta \{\stackrel{n}{\stackunder{k=1}{\sum }}(2a_{k}^{s}I_{kz})+%
\stackrel{n}{\stackunder{1=k<l<m}{\sum }}%
(2a_{k}^{s}I_{kz})(2a_{l}^{s}I_{lz})(2a_{m}^{s}I_{mz})+...\}]\ $ $\quad (39)$%
\newline
where the effective Hamiltonian $H_{ss}$ is written as

$\theta H_{ss}=N\theta E_{ss}-\exp (-i\pi F_{y})N\theta E_{ss}\exp (i\pi
F_{y})\qquad \qquad \qquad $

$=2\theta \{\stackrel{n}{\stackunder{k=1}{\sum }}(2a_{k}^{s}I_{kz})+%
\stackrel{n}{\stackunder{1=k<l<m}{\sum }}%
(2a_{k}^{s}I_{kz})(2a_{l}^{s}I_{lz})(2a_{m}^{s}I_{mz})+...\}\ \qquad \qquad
(40)$ \newline
where $N=2^{n}.$ Since one can manipulate at will any individual two-state
particle of the quantum system all the interactions involving only the $j$th
two-state particle $(j=1,2,...,n)$ on the right-hand side of Eq.(39) then
can be extracted by using the spin-echo sequence:

$\ U_{sj}=U_{ss}\exp (-i\pi I_{jy})U_{ss}^{+}\exp (i\pi I_{jy})$

$\qquad =\exp [-i4\theta (2a_{j}^{s}I_{jz})\{1+\stackrel{n}{\stackunder{1=k<l%
}{\sum^{\prime }}}(2a_{k}^{s}I_{kz})(2a_{l}^{s}I_{lz})+...\}]\ \qquad \qquad
\quad (41)$ \newline
where the sums $\sum^{\prime }$ with prime symbol run over all indexes
except the index $j$. There are only even-body interactions in addition to
the unity operator in the bracket \{\} on the right-hand side of Eq.(41). In
order to cancel further all these even-body interactions but leave only the
unity operator in the bracket \{\} one may first make a simple unitary
transformation on the unitary operator $U_{sj}$ to convert all these
even-body interactions into multiple-quantum coherence operators and then
uses the phase cycling techniques [23, 27] to cancel them (see next
sections),

$\exp (-i\theta H_{jQ})\equiv \exp (-i\frac{\pi }{2}F_{jy})U_{sj}\exp (i%
\frac{\pi }{2}F_{jy})$

$\qquad \qquad \quad \ \ =\exp (-i4\theta (2a_{j}^{s}I_{jz})\{1+\stackrel{n}{%
\stackunder{1=k<l}{\sum^{\prime }}}(2a_{k}^{s}I_{kx})(2a_{l}^{s}I_{lx})$

$\qquad \qquad +\stackrel{n}{\stackunder{1=k<l<p<q}{\sum^{\prime }}}%
(2a_{k}^{s}I_{kx})(2a_{l}^{s}I_{lx})(2a_{p}^{s}I_{px})(2a_{q}^{s}I_{qx})+...%
\})$ \qquad \qquad $(42)$\newline
where the effective Hamiltonian $H_{jQ}$ is given by

$\theta H_{jQ}=\exp (-i\frac{\pi }{2}F_{jy})\{\theta H_{ss}$

$\qquad \qquad \qquad -\exp (-i\pi I_{jy})\theta H_{ss}\exp (i\pi
I_{jy})\}\exp (i\frac{\pi }{2}F_{jy})\qquad \qquad \quad \ \ (43)$ \newline
and it satisfies

$\theta H_{jQ}\equiv \exp (-i\frac{\pi }{2}F_{jy})\{N\theta E_{ss}$

$\qquad -\exp (-i\pi F_{y})N\theta E_{ss}\exp (i\pi F_{y})\}\exp (i\frac{\pi 
}{2}F_{jy})$

$\qquad -\exp (-i\frac{\pi }{2}F_{jy})\exp (-i\pi I_{jy})\{N\theta E_{ss}$

$\qquad -\exp (-i\pi F_{y})N\theta E_{ss}\exp (i\pi F_{y})\}\exp (i\pi
I_{jy})\exp (i\frac{\pi }{2}F_{jy})$

$=$ $4\theta (2a_{j}^{s}I_{jz})\{1+\stackrel{n}{\stackunder{1=k<l}{%
\sum^{\prime }}}(2a_{k}^{s}I_{kx})(2a_{l}^{s}I_{lx})$

$\qquad +\stackrel{n}{\stackunder{1=k<l<p<q}{\sum^{\prime }}}%
(2a_{k}^{s}I_{kx})(2a_{l}^{s}I_{lx})(2a_{p}^{s}I_{px})(2a_{q}^{s}I_{qx})+...%
\}\qquad \qquad \quad \quad \ (44)$ \newline
where the operator $F_{jp}=$ $\stackrel{n}{\stackunder{k=1,k\neq j}{\sum }}%
I_{kp}$ $(p=x,y,z).$ Note that there are the identities for the selective
phase-shift operation $C_{s}(\theta )$:

$C_{s}(\pm N\theta )=\exp (\mp i\theta 2^{n}E_{ss})$

$\qquad \qquad \ =\exp [\mp i(2^{n}\theta \func{mod}(2\pi
))E_{ss}]=C_{s}(\pm N\theta \func{mod}(2\pi )).$ $\quad \qquad \ \ (45)$ 
\newline
It follows from Eqs.(41)-(44) that the unitary operator $\exp (-i\theta
H_{jQ})$ can be expressed explicitly as

$\exp (-i\theta H_{jQ})=\exp (-i\theta H_{jq})$

$=\exp (-i\frac{\pi }{2}F_{jy})C_{s}(N\theta \func{mod}(2\pi ))\exp (-i\pi
F_{y})C_{s}(-N\theta \func{mod}(2\pi ))$

$\quad \times \exp (-i\pi I_{jy})C_{s}(N\theta \func{mod}(2\pi ))\exp (i\pi
F_{y})C_{s}(-N\theta \func{mod}(2\pi ))$

$\quad \times \exp (i\pi I_{jy})\exp (i\frac{\pi }{2}F_{jy})\ \quad \qquad
\qquad \qquad \qquad \qquad \qquad \quad \qquad \qquad \ (46)$ \newline
where the effective Hamiltonian $H_{jq}$ is written as

$\theta H_{jq}=\exp (-i\frac{\pi }{2}F_{jy})\{N\theta \func{mod}(2\pi
)E_{ss} $

$\qquad -\exp (-i\pi F_{y})N\theta \func{mod}(2\pi )E_{ss}\exp (i\pi
F_{y})\}\exp (i\frac{\pi }{2}F_{jy})$

$\qquad -\exp (-i\frac{\pi }{2}F_{jy})\exp (-i\pi I_{jy})\{N\theta \func{mod}%
(2\pi )E_{ss}\qquad $\newline
$-\exp (-i\pi F_{y})N\theta \func{mod}(2\pi )E_{ss}\exp (i\pi F_{y})\}\exp
(i\pi I_{jy})\exp (i\frac{\pi }{2}F_{jy}).\qquad \quad \ \ (47)$

One can see from Eqs.(44) and (47) that the effective Hamiltonian $\theta
H_{jQ}$ is proportional to $N\theta $ but the Hamiltonian $\theta H_{jq}$ is
bounded by the norm $\left\| \theta H_{jq}\right\| $ $\leq 8\pi $ for any
qubit number $n.$ It is well known that the highest order of
multiple-quantum coherence is $n$ for a coupled spin (I=1/2) system with $n$
two-state particles [23]. This indicates that the operator $H_{jQ}$ of
Eq.(44) consists of a variety of multiple-quantum coherence operators whose
quantum orders take any values from $-(n-1)$ to $(n-1)$ due to the fact that
the operator $(2a_{j}^{s}I_{jz})$ in the operator $H_{jQ}$ is the $LOMSO$
operator, which is also a zero-quantum operator [18]. To cancel all these
multiple-quantum coherence operators but leave only the desired term $%
(2a_{j}^{s}I_{jz})$ on the right-hand side of the operator $H_{jQ}$ of
Eq.(44) the discrete Fourier analysis [28] and the phase cycling technique
[23, 27] will be exploited below. \newline
\newline
\textbf{7.1 The discrete Fourier analysis and the phase cycling technique}

The phase cycling technique is one of the most useful methods in nuclear
magnetic resonance (NMR) spectroscopy [23, 27]. It has been used extensively
to select the specific multiple-quantum coherences with the desired quantum
order and cancel any other undesired multiple-quantum coherences in the NMR
experiments. In principle, the phase cycling technique is based on the
discrete Fourier transform [28]. The principle of the phase cycling
technique can be outlined below. Consider a coupled spin (I=1/2) system with 
$n$ nuclear spins. The density operator $\rho (t)$ of the system then can be
classified generally according to different quantum order $p$

$\qquad \qquad \qquad \qquad \rho (t)=\stackunder{p=-n}{\stackrel{n}{\sum }}$
$\rho ^{p}(t).\qquad \qquad \qquad \qquad \qquad \qquad \qquad \ \ (48)$%
\newline
A $p$-order quantum coherence has an important characteristic transformation
behavior under the z-axis rotations [23, 27]

$\qquad \qquad \exp (-i\varphi F_{z})\rho ^{p}(t)\exp (i\varphi F_{z})=\rho
^{p}(t)\exp (-ip\varphi ).\qquad \qquad \quad \ \ (49)$ \newline
That is, a $p$-order quantum coherence generates a phase shift proportional
to its quantum order $p$ under the z-axis rotations. This key property is
the base to separate different order quantum coherences, to cancel the
undesired order and to select the desired order quantum coherences in a
coupled multi-spin (I=1/2) system. Now by making a series of the z-axis
rotations with systematic increments of the phase angle $\varphi _{k}$ on
the $p$-order quantum coherence,

$\qquad \qquad \qquad \varphi _{k}=k2\pi /N,\quad k=0,1,2,...,N-1,$ \newline
and then summing up all the rotational results one obtains

$\frac{1}{N}\stackrel{N-1}{\stackunder{k=0}{\sum }}\exp (-i\varphi
_{k}F_{z})\rho ^{p}(t)\exp (i\varphi _{k}F_{z})$

$\qquad \qquad =\rho ^{p}(t)\{\frac{1}{N}\stackrel{N-1}{\stackunder{k=0}{%
\sum }}\exp (-i2\pi pk/N)\}.\qquad \qquad \qquad \qquad \qquad \quad \ \
(50) $ \newline
Note that there is the exponential sum relation in the conventional discrete
Fourier transform [28],

$\qquad \frac{1}{N}\stackrel{N-1}{\stackunder{k=0}{\sum }}\exp (-i2\pi
pk/N)=\{ 
\begin{array}{l}
1,\qquad p=0 \\ 
0,\qquad p\neq 0
\end{array}
\qquad \qquad \qquad \qquad \qquad (51)$\newline
where $p$ is an integer and $N>|p|$. One can reduce the sum of Eq.(50) to
the form

$\frac{1}{N}\stackrel{N-1}{\stackunder{k=0}{\sum }}\exp (-i\varphi
_{k}F_{z})\rho ^{p}(t)\exp (i\varphi _{k}F_{z})=\{ 
\begin{array}{l}
\rho ^{0}(t),\qquad p=0 \\ 
0,\qquad \quad \ \ p\neq 0
\end{array}
\qquad \qquad \ \ (52)$ \newline
Now applying the phase cycling technique to the density operator of Eq.(48)
one obtains with the help of Eq.(52)

$\frac{1}{N}\stackrel{N-1}{\stackunder{k=0}{\sum }}\exp (-i\varphi _{k}F_{z})%
\stackunder{p=-n}{\stackrel{n}{\sum }}\rho ^{p}(t)\exp (i\varphi
_{k}F_{z})=\rho ^{0}(t)\qquad \qquad \qquad \qquad \ \ (53)$ \newline
where $N>n$. Equation (53) shows that by $N$-step phase cycling
systematically only the zero-quantum coherence components ($p=0$) in the
density operator of Eq.(48) keep unchanged but all other nonzero order
quantum coherences ($p\neq 0$) are cancelled.

Because the operator $H_{jQ}$ of Eq.(44) consists of multiple-quantum
coherences with different quantum orders, by replacing the density operator
in Eq.(53) with the operator $H_{jQ}$ and with the help of the phase cycling
technique all the $p$-order quantum coherences $(p\neq 0)$ are cancelled in
the operator $H_{jQ}$ and one will obtain only the zero-quantum coherences

$\qquad \frac{1}{N}\stackrel{N-1}{\stackunder{k=0}{\sum }}\exp (-i\varphi
_{jk}F_{jz})H_{jQ}\exp (i\varphi _{jk}F_{jz})=H_{jQ}^{0}\qquad \qquad \qquad
\qquad \ (54)$ \newline
where $N>n-1$. Note that the operators $I_{kx}=\frac{1}{2}%
(I_{k}^{+}+I_{k}^{-})$ and $I_{ky}=\frac{1}{2i}(I_{k}^{+}-I_{k}^{-}).$ It
turns out easily from Eq.(44) and (54) that the zero-quantum operator $%
H_{jQ}^{0}$ can be expressed explicitly as

$H_{jQ}^{0}=4(2a_{j}^{s}I_{jz})\{1+\stackrel{n}{\stackunder{k\neq l}{%
\sum^{\prime }}}J_{k,l}(I_{k}^{+}I_{l}^{-})$

$\qquad \qquad \qquad +\stackrel{n}{\stackunder{k\neq l\neq p\neq q}{%
\sum^{\prime }}}J_{kl,pq}(I_{k}^{+}I_{l}^{+}I_{p}^{-}I_{q}^{-})+...\}\ \
\qquad \qquad \qquad \qquad \quad \ (55)$ \newline
where the coefficients $J_{k,l}$, $J_{kl,pq}$, etc., depend on the
unity-number vector $\{a_{k}^{s}\}.$ The Hermiticity of the operator $%
H_{jQ}^{0}$ shows that the coefficients satisfy $J_{k,l}=J_{l,k}^{*}$, $%
J_{kl,pq}=J_{pq,kl}^{*}$, etc.. Equation (54) shows that by $N=n$ step phase
cycling one can cancel all the nonzero-order multiple-quantum coherences in
the operator $H_{jQ}.$ However, equations (54) and (55) also show that the
zero-quantum coherences in the operator $H_{jQ}$ are not cancelled by the
phase cycling technique. Therefore, one needs to cancel further the residual
zero-quantum coherences in the zero-quantum operator $H_{jQ}^{0}$ so as to
obtain the desired longitudinl magnetizaton term $(2a_{j}^{s}I_{jz}).$%
\newline
\newline
\textbf{7.2 The cancellation for the zero-quantum coherences}

There is also the important characteristic transformation behavior under the
z-axis rotations applied to each individual spin in a spin system, like
Eq.(49) [23],

$\qquad \exp (-i2\pi pf_{k}I_{kz})I_{k}^{\pm }\exp (i2\pi
pf_{k}I_{kz})=I_{k}^{\pm }\exp (\mp i2\pi pf_{k}),\qquad \qquad (56)$ 
\newline
where $f_{k}$ is called the offset frequency of the $k$th spin in the spin
system. Note that one can manipulate each individual spin in the system by
an external field. By making a series of the z-axis rotations applied to
each individual spin on the zero-quantum operator $H_{jQ}^{0}$ and then
summing up all the rotational results one obtains from Eq.(55) with the help
of Eq.(56)

$\frac{1}{N}\stackrel{N-1}{\stackunder{m=0}{\sum }}\exp [-i(2\pi m/N)%
\stackrel{n}{\stackunder{k=1,k\neq j}{\sum }}f_{k}I_{kz}]H_{jQ}^{0}\exp
[i(2\pi m/N)\stackrel{n}{\stackunder{k=1,k\neq j}{\sum }}f_{k}I_{kz}]=$

$4(2a_{j}^{s}I_{jz})\{1+\stackrel{n}{\stackunder{k\neq l}{\sum^{\prime }}}%
\{J_{k,l}(I_{k}^{+}I_{l}^{-})\frac{1}{N}\stackrel{N-1}{\stackunder{m=0}{\sum 
}}\exp [-i(2\pi m/N)(f_{k}-f_{l})]\}$

$+\stackrel{n}{\stackunder{k\neq l\neq p\neq q}{\sum^{\prime }}}%
\{J_{kl,pq}(I_{k}^{+}I_{l}^{+}I_{p}^{-}I_{q}^{-})$

$\times \frac{1}{N}\stackrel{N-1}{\stackunder{m=0}{\sum }}\exp [-i(2\pi
m/N)(f_{k}+f_{l}-f_{p}-f_{q})]\}+...\}.\qquad \qquad \qquad \qquad (57)$ 
\newline
If all the frequencies of the zero-quantum coherences such as $f_{k}-f_{l}$, 
$f_{k}+f_{l}-f_{p}-f_{q}$, etc. are taken as nonzero integer, it follows
from the exponential sum relation (51) that all the zero-quantum coherences
on the right-hand side of Eq.(57) are cancelled exactly, leaving only the
desired term $4(2a_{j}^{s}I_{jz})$ when the step number $N$ is larger than
the maximum zero-quantum frequency in Eq.(57). How to choose explicitly the
proper offset frequency set $\{f_{1},$ $f_{2},$ $..,$ $f_{n}\}$ so that all
the zero-quantum coherences in Eq.(57) can be cancelled? The suitable
integer set of $\{f_{1},$ $f_{2},$ $..,$ $f_{n}\}$ should satisfy that (a)
all the zero-quantum frequencies such as $f_{k}-f_{l}$, $%
f_{k}+f_{l}-f_{p}-f_{q}$, etc., take nonzero integers and (b) there is a
minimum step number $N<poly(n)$ that does not divide each zero-quantum
frequency. The condition $N<poly(n)$ ensures that all the zero-quantum
coherences in the operator $H_{jQ}^{0}$ can be cancelled by the phase
cycling of Eq.(57) with a polynomial step number $N$. It is easy to find an
offset frequency set $\{f_{k}\}$ that satisfies only the condition (a). As
an example, all the zero-quantum frequencies take nonzero integers if the
offset frequency set $\{f_{k}\}$ is taken as a super-ascend integer series: $%
f_{k}>0$ and $f_{k+1}>\stackrel{k}{\stackunder{l=1}{\sum }}f_{l},$ e.g., $%
\{f_{k}\}=\{2^{0},2^{1},...,2^{n-1}\}.$ However, it may be really difficult
to find the minimum step number $N$ so that $N<poly(n)$ when the condition
(a) is met$.$ Without losing generality, assume below that all the
zero-quantum frequencies $f_{zq}$ take nonzero values. With the aid of
Eq.(56) the following integration identity is constructed similar to the
summation of Eq.(57)

$\frac{1}{2T}\stackrel{T}{\stackunder{-T}{\int }}dt\exp [-i2\pi t\stackrel{n%
}{\stackunder{k=1,k\neq j}{\sum }}f_{k}I_{kz}]H_{jQ}^{0}\exp [i2\pi t%
\stackrel{n}{\stackunder{k=1,k\neq j}{\sum }}f_{k}I_{kz}]=$

$4(2a_{j}^{s}I_{jz})\{1+\stackrel{n}{\stackunder{k\neq l}{\sum^{\prime }}}%
\{J_{k,l}(I_{k}^{+}I_{l}^{-})\frac{1}{2T}\stackrel{T}{\stackunder{-T}{\int }}%
dt\exp [-i(2\pi t(f_{k}-f_{l})]\}$

$+\stackrel{n}{\stackunder{k\neq l\neq p\neq q}{\sum^{\prime }}}%
\{J_{kl,pq}(I_{k}^{+}I_{l}^{+}I_{p}^{-}I_{q}^{-})$

$\times \frac{1}{2T}\stackrel{T}{\stackunder{-T}{\int }}dt\exp [-i2\pi
t(f_{k}+f_{l}-f_{p}-f_{q})]\}+...\}.\qquad \qquad \qquad \qquad \qquad \quad
\ (58)$ $\newline
$Note that for a sufficiently large positive number $T$ and any nonzero
zero-quantum frequency $f_{zq},$\qquad

$\qquad \frac{1}{2T}\stackrel{T}{\stackunder{-T}{\int }}dt\exp (-i2\pi
tf_{zq})=\dfrac{\sin (2\pi Tf_{zq})}{2\pi Tf_{zq}}=0.\qquad \qquad \qquad
\qquad \quad \ (59)$ \newline
Inserting Eq.(59) into Eq.(58) and making a variable substitution $t=\lambda
T$ one obtains

$4(2a_{j}^{s}I_{jz})=\frac{1}{2}\stackrel{1}{\stackunder{0}{\int }}d\lambda
\{\exp [-i2\pi \lambda T\stackrel{n}{\stackunder{k=1,k\neq j}{\sum }}%
f_{k}I_{kz}]$

$\qquad \times H_{jQ}^{0}\exp [i2\pi \lambda T\stackrel{n}{\stackunder{%
k=1,k\neq j}{\sum }}f_{k}I_{kz}]$

$+\exp [i2\pi \lambda T\stackrel{n}{\stackunder{k=1,k\neq j}{\sum }}%
f_{k}I_{kz}]H_{jQ}^{0}\exp [-i2\pi \lambda T\stackrel{n}{\stackunder{%
k=1,k\neq j}{\sum }}f_{k}I_{kz}]\}.\qquad \qquad \ \ (60)$ \newline
The oracle unitary operation exp$(-i\theta a_{j}^{s}I_{jz})$ can be
conveniently expressed as, by inserting the zero-quantum operator $%
H_{jQ}^{0} $ of Eq.(54) into Eq.(60),

$\exp (-i\theta a_{j}^{s}I_{jz})=\exp \{-i\frac{\theta }{16n}\stackrel{n-1}{%
\stackunder{k^{\prime }=0}{\sum }}\exp (-i\varphi _{jk^{\prime
}}F_{jz})(H_{jQ}(T)$

$\qquad \qquad \qquad \qquad \qquad +H_{jQ}(-T))\exp (i\varphi _{jk^{\prime
}}F_{jz})\}\qquad \qquad \qquad \quad \ \quad (61)$ \newline
where the Hermitian operator $H_{jQ}(T)$ is written as

$H_{jQ}(T)=\stackrel{1}{\stackunder{0}{\int }}d\lambda \{\exp (-i2\pi
\lambda T\stackrel{n}{\stackunder{k=1,k\neq j}{\sum }}f_{k}I_{kz})$

$\qquad \qquad \times H_{jQ}\exp (i2\pi \lambda T\stackrel{n}{\stackunder{%
k=1,k\neq j}{\sum }}f_{k}I_{kz})\}\ \qquad \qquad \qquad \qquad \qquad \quad
\ \ \ (62)$ \newline
with the matrix element of any pair of the conventional computational base $%
|r\rangle $ and $|t\rangle :$

$\langle r|H_{jQ}(T)|t\rangle =\stackrel{1}{\stackunder{0}{\int }}d\lambda
\langle r|H_{jQ}|t\rangle \exp [-i\lambda T\pi \stackrel{n}{\stackunder{%
k=1,k\neq j}{\sum }}(a_{k}^{r}-a_{k}^{t})f_{k}].$\newline
Obviously, $\langle r|H_{jQ}(T)|t\rangle =\langle r|H_{jQ}|t\rangle $ if the
multiple-quantum transition frequency equals zero, that is, $\stackrel{n}{%
\stackunder{k=1,k\neq j}{\sum }}(a_{k}^{r}-a_{k}^{t})f_{k}=0;$ otherwise, $%
\langle r|H_{jQ}(T)|t\rangle =0.$ As assumed previously, the zero-quantum
transition frequency $\stackrel{n}{\stackunder{k=1,k\neq j}{\sum }}%
(a_{k}^{r}-a_{k}^{t})f_{k}\neq 0$ where the transition of the pair of the
computational base $|r\rangle $ and $|t\rangle $ is a zero-quantum
transition. Then, $\langle r|H_{jQ}(T)|t\rangle =0$ for any zero-quantum
transition betwwen the base $|r\rangle $ and $|t\rangle $. For convenient
discussion, first of all, assume that all multiple-quantum transition
frequencies including the zero-quantum transition frequencies take nonzero
values. Then the transition frequencies $T[\stackrel{n}{\stackunder{%
k=1,k\neq j}{\sum }}(a_{k}^{r}-a_{k}^{t})f_{k}]$ are large numbers, that is, 
$\left| Tf_{k}\right| \gg 1,$ for a sufficiently large number $T$,
indicating that the integral of Eq.(62) contains rapidly oscillating
periodic integrand. The conventional numerical integration method may not be
available for the type of rapidly oscillating integrals [29, 30]. This
one-dimensional rapidly oscillating integral may be first converted into a
multiple integral before integrating it numerically and this usually will
generate an error of $O(\frac{1}{T})$ [30]. It turns out easily that this
error is actually zero for the rapidly oscillating periodic integrand
operator of Eq.(62) when $T$ is a sufficiently large number. Let $%
y_{k}=\lambda Tf_{k},$ $k=1,2,...,n.$ The one-dimensional rapidly
oscillating periodic integral (62) then is converted exactly into the
multiple integral:

$H_{jQ}(T)=\stackrel{1}{\stackunder{0}{\int }}...\stackrel{1}{\stackunder{0}{%
\int }}dy_{1}dy_{2}...dy_{n}\{\exp (-i2\pi \stackrel{n}{\stackunder{%
k=1,k\neq j}{\sum }}y_{k}I_{kz})$

$\qquad \qquad \qquad \qquad \times H_{jQ}\exp (i2\pi \stackrel{n}{%
\stackunder{k=1,k\neq j}{\sum }}y_{k}I_{kz})\}.\qquad \qquad \qquad \qquad
\quad \ (63)$ \newline
In fact, the equal matrix element $[H_{jQ}(T)]_{kl}$ of any pair of the
usual computational bases $|k\rangle $ and $|l\rangle $ can be obtained from
the multiple integral (63) and from the one-dimensional integral (62),
respectively. It easily proves from Eqs.(44) and (62) that the Hermitian
operator $H_{jQ}(T)$ is a diagonal operator with the norm $\left\|
H_{jQ}(T)\right\| =4$ and commutes with the diagonal unitary operators $\exp
(\pm i\varphi _{jk}F_{jz}).$ Therefore, equation (61) is reduced to the
simple form

$\exp (-i\theta a_{j}^{s}I_{jz})=\exp \{-i\frac{1}{16}\theta H_{jQ}(T)\}\exp
\{-i\frac{1}{16}\theta H_{jQ}(-T)\}.\qquad \quad \quad \ \ (64)$ \newline
This result shows that in the case that all multiple-quantum (including
zero-quantum) transition frequencies take nonzero values the phase cycling
in section 7.1 to cancel all the nonzero-order multiple quantum coherences
becomes not necessary. There are a number of numerical integration methods
to calculate a multiple integral [30-36]. One simple numerical method is
Hua-Wang method [31, 32] based on the number theory [37]. The lattice point
for the numerical multiple integration is chosen as a real algebraic
irrational point $(y_{1},y_{2},...,y_{n})=\mathbf{\omega }=(\omega
_{1},\omega _{2},...,\omega _{n}),$ where $\{1,\omega _{1,}\omega
_{2},...,\omega _{n}\}$ are linearly independent real algebraic numbers over
the rationals$.$ Then the multiple integral (63) can be replaced with a
discrete summation through the numerical multidimensional integration [31,
32]

$\stackrel{1}{\stackunder{0}{\int }}...\stackrel{1}{\stackunder{0}{\int }}%
dy_{1}dy_{2}...dy_{n}G_{jQ}(y_{1},y_{2},...,y_{n})$

$=\dfrac{1}{(2M+1)^{l}}\stackrel{M\times l}{\stackunder{k=-M\times l}{\sum }}%
\Phi (M,l,k)G_{jQ}(k\omega _{1},k\omega _{2},...,k\omega _{n})+O(M,l)\quad
\quad \ (65)$ \newline
where the integrand operator is given by

$G_{jQ}(y_{1},y_{2},...,y_{n})$

$\qquad =\exp (-i2\pi \stackrel{n}{\stackunder{k=1,k\neq j}{\sum }}%
y_{k}I_{kz})H_{jQ}\exp (i2\pi \stackrel{n}{\stackunder{k=1,k\neq j}{\sum }}%
y_{k}I_{kz})$\ \qquad $\qquad \quad \ (66)$ \newline
with the matrix element of any pair of the computational base $|r\rangle $
and $|t\rangle :$

$\langle r|G_{jQ}(y_{1},y_{2},...,y_{n})|t\rangle =\langle r|H_{jQ}|t\rangle
\exp [-i\pi \stackrel{n}{\stackunder{k=1,k\neq j}{\sum }}%
(a_{k}^{r}-a_{k}^{t})y_{k}].$\qquad $\qquad \quad $\newline
The integer weight distributions $\Phi (M,l,k)$ in Eq.(65) are determined
from the polynomial identity

$\qquad \qquad (\stackrel{M}{\stackunder{k=-M}{\sum }}z^{k})^{l}=\stackrel{%
M\times l}{\stackunder{k=-M\times l}{\sum }}\Phi (M,l,k)z^{k},\qquad \qquad
\qquad \qquad \qquad \quad \ (67)$ \newline
and the error function operator $O(M,l)$ can be derived as [31, 32] (also
see Appendix B)

$\langle r|O(M,l)|t\rangle =\{ 
\begin{array}{l}
-\langle r|H_{jQ}|t\rangle (\dfrac{\sin [(2M+1)\pi (\mathbf{m}(r,t),\mathbf{%
\omega })]}{(2M+1)\sin [\pi (\mathbf{m}(r,t),\mathbf{\omega })]})^{l},\ \
r\neq t\quad \\ 
0,\quad r=t
\end{array}
(68)$ \newline
where the nonzero integer vector $\mathbf{m}%
(r,t)=(m_{1}(r,t),m_{2}(r,t),...,m_{n}(r,t))$ with $%
m_{k}(r,t)=(a_{k}^{r}-a_{k}^{t})/2=-1,0,+1;$ $k=1,2,...,n.$ The inner
product between the two vectors $\mathbf{m}(r,t)$ and $\mathbf{\omega }$ is
defined as $(\mathbf{m}(r,t),\mathbf{\omega })=\stackrel{n}{\stackunder{k=1}{%
\sum }}\omega _{k}m_{k}(r,t).$ By using the Chebyshev polynomial of the
second kind $U_{n}($cos$\varphi )=\sin ((n+1)\varphi )/\sin (\varphi )$ the
error function of Eq.(68) is reduced to the form

$\langle r|O(M,l)|t\rangle =-\langle r|H_{jQ}|t\rangle [\dfrac{U_{2M}(\cos
\varphi )}{(2M+1)}]^{l},\quad r\neq t\qquad \qquad \qquad \qquad \quad (69)$%
\newline
where $\left| \dfrac{U_{2M}(\cos \varphi )}{(2M+1)}\right| \leq 1,$ $\varphi
=\varphi (r,t)=\pi \langle (\mathbf{m}(r,t),\mathbf{\omega })\rangle $, $%
\langle x\rangle $ denotes the distance from the real number $x$ to the
nearest integer, i.e, $\left\langle x\right\rangle =\min (\left\{ x\right\}
,1-\left\{ x\right\} ),$ and $\left\{ x\right\} $ is the decimal part of the
real $x$. By Eq.(44) one can prove that $\left| \langle r|H_{jQ}|t\rangle
\right| =4$ for any pair of the computational base $|r\rangle $ and $%
|t\rangle .$ Then it follows from Eqs.(68) and (69) that the error function
operator $O(M,l)$ of Eq.(65) is bounded by

$\left\| O(M,l)\right\| \leq [\stackrel{2^{n}-1}{\stackunder{r=0}{\sum }}%
\stackrel{2^{n}-1}{\stackunder{t=0}{\sum }}\left| \langle r|O(M,l)|t\rangle
\right| ^{2}]^{1/2}$

$\qquad \qquad \qquad \leq \dfrac{4}{(2M+1)^{l}}[\stackrel{2^{n}-1}{%
\stackunder{r,t=0,r\neq t}{\sum }}\left| U_{2M}(\cos \varphi (r,t))\right|
^{2l}]^{1/2}.\qquad \quad \qquad (70)$ \newline
This error function can be made as small as desired when the real algebraic
irrational lattice point $\mathbf{\omega }$ and the numbers $M$ and $l$ in
the numerical integration (65) are chosen suitably. The explicit estimate
for the error upper bound may be achieved with the Hua-Wang numerical method
[31, 32]. As shown in Refs.[31-35, 38], for every real algebraic irrational
point $\mathbf{\omega }$ and a finite positive number $a\geq 1$ there exists
a positive constant $b=b(a,\mathbf{\omega })>0$ dependent only on the vector 
$\mathbf{\omega }$ and the number $a$ such that

$\qquad \quad \qquad \left\langle (\mathbf{m}(r,t),\mathbf{\omega }%
)\right\rangle \geq b(a,\mathbf{\omega })h\mathbf{(m}(r,t))^{-a}\qquad
\qquad \qquad \qquad \quad \ \ (71)$ \newline
holds for all nonzero integer vectors $\mathbf{m}(r,t)$ (the Schmidt theorem
[38])$.$ In the Schmidt theorem (71), the distance of the integer vector $%
\mathbf{m}(r,t)$ from the origin is defined by $h(\mathbf{m}(r,t))=\stackrel{%
n}{\stackunder{k=1}{\prod }}\max (1,\left| m_{k}(r,t)\right| )$. Therefore, $%
h(\mathbf{m}(r,t))=1$ for any computational base $|r\rangle $ and $|t\rangle 
$ in the case of the integrand operator of Eq.(66). Note that $\left| \sin
[\pi (\mathbf{m}(r,t),\mathbf{\omega })]\right| \geq 2\left\langle (\mathbf{m%
}(r,t),\mathbf{\omega })\right\rangle $ [31-35]$.$ From the inequalities
(70) and (71) it follows that the error function operator is bounded by

$\qquad \quad \left\| O(M,l)\right\| \leq 4(4^{n}-2^{n})^{1/2}(\dfrac{1}{%
2(2M+1)b(a,\mathbf{\omega })})^{l}.\qquad \qquad \qquad \quad (72)$ \newline
By choosing suitably the numbers $M$ and $l,$ for example, $l\sim n$, $%
2M+1\sim b(a,\mathbf{\omega })^{-1},$ one can get the desired small error
function operator $O(M,l)$ in the numerical integration (65). It is believed
that the coefficient $b(a,\mathbf{\omega })>0$ does not decrease
exponentially as the dimension size $n$ of the multiple integral (63) since
it is dependent only on both the algebraic irrational $\mathbf{\omega }$ and
the number $a$ [31-35, 38]. As a consequence, the lattice point number $%
(2M\times l+1)$ in the numerical integration (65) does not grow
exponentially as the dimension size $n$.

The Schmidt theorem (71) is still too strong for the numerical multiple
integration (65), although the integration (65) requires that the Schmidt
theorem (71) hold only for all those nonzero integer vectors $\mathbf{m}%
(r,t) $ that satisfy

$\qquad \qquad \qquad \mathbf{m}(r,t)\neq 0,\ h(\mathbf{m}(r,t))=1\qquad
\qquad \qquad \qquad \qquad \ \ \qquad (73)$ \newline
instead of for all the nonzero integer vectors $\mathbf{m}(r,t)$. In fact,
the error upper bounds (70) and (72) are derived under the previous
assumption that all the multiple-quantum transition frequencies take nonzero
values. This assumption really corresponds to the condition (73). One can
weaken the use of the Schmidt theorem (71) in the evaluation of the error
function (70) with the help of the phase cycling technique in section 7.1.
Provided that the phase cycling is available in Eq.(61), the Schmidt theorem
(71) is required to hold only for those nonzero integer vectors $\mathbf{m}%
(r,t)$ that satisfy

$\qquad \qquad \mathbf{m}(r,t)\neq 0,\ h(\mathbf{m}(r,t))=1,$ $\stackrel{n}{%
\stackunder{k=1}{\sum }}m_{k}(r,t)=0.\qquad \qquad \qquad \ \ (74)$ \newline
Obviously, the assumption that all the zero-quantum transition frequencies
take nonzero values corresponds to the condition (74). Although those
nonzero integer vectors $\mathbf{m}(r,t)$ that do not satisfy the condition
(74) can cause a large error on the right-hand side of Eq.(65), the error is
of the nonzero-order multiple-quantum coherences and hence can be further
cancelled by the phase cycling technique, as shown in section 7.1. Thus,
inserting the numerical integration of Eq.(65) into Eq.(61) one obtains

$\exp (-i\theta a_{j}^{s}I_{jz})=\exp \{-i\frac{\theta }{16n}\stackrel{n-1}{%
\stackunder{k^{\prime }=0}{\sum }}\exp (-i\varphi _{jk^{\prime }}F_{jz})$

$\times \stackrel{M\times l}{\stackunder{k=-M\times l}{\sum }}\dfrac{\Phi
(M,l,k)}{(2M+1)^{l}}[G_{jQ}(k\omega _{1},k\omega _{2},...,k\omega _{n})$

$+G_{jQ}(-k\omega _{1},-k\omega _{2},...,-k\omega _{n})]\exp (i\varphi
_{jk^{\prime }}F_{jz})-iO_{0}(M,l)\}\qquad \qquad \quad \ \ (75)$ \newline
where the error function operator is given by

$O_{0}(M,l)=\frac{\theta }{8n}\stackrel{n-1}{\stackunder{k^{\prime }=0}{\sum 
}}\exp (-i\varphi _{jk^{\prime }}F_{jz})O(M,l)\exp (i\varphi _{jk^{\prime
}}F_{jz})\qquad \quad \qquad \ \ \ (76)$ \newline
with the error function operator $O(M,l)$ given by Eq.(68) or (69).
Obviously, $\langle r|O_{0}(M,l)|t\rangle =0$ for all the nonzero integer
vectors $m(r,t)$ that do not satisfy the condition (74)$.$ Therefore, one has

$\langle r|O_{0}(M,l)|t\rangle =\{ 
\begin{array}{l}
\frac{\theta }{8}\langle r|O(M,l)|t\rangle ,\text{ the conditions (74) hold}
\\ 
0,\text{ otherwise}
\end{array}
\qquad \qquad (77)$ \newline
and the error upper bound is derived as

$\left\| O_{0}(M,l)\right\| \leq \dfrac{4}{(2M+1)^{l}}[\stackrel{2^{n}-1}{%
\stackunder{r,t=0,r\neq t}{\sum^{\prime }}}\left| U_{2M}(\cos \varphi
(r,t))\right| ^{2l}]^{1/2}\qquad \quad \ \qquad \ (78a)$ \newline
where the sums run only over all pair of the computational base $|r\rangle $
and $|t\rangle $ of the zero-quantum transition. It follows from (68)-(71)
and Eq.(77) that the error function operator $O_{0}(M,l)$ has the explicit
upper bound:

$\left\| O_{0}(M,l)\right\| \leq \frac{1}{2}\left| \theta \right| (( 
\begin{array}{l}
2n \\ 
n
\end{array}
)-2^{n})^{1/2}(\dfrac{1}{2(2M+1)b(a,\mathbf{\omega })})^{l}\qquad \qquad
\qquad (78b)$ \newline
because number of all the linearly independent zero-quantum coherences in
the quantum system with $n$ two-state particles is given by [23]

$\qquad \qquad Z_{0}=( 
\begin{array}{l}
2n \\ 
n
\end{array}
)-2^{n}.$\newline
Obviously, the error upper bound $\left\| O_{0}(M,l)\right\| $ is much
smaller than $\left\| O(M,l)\right\| $.

How to express the exponential operator on the right-hand side of Eq.(64) or
Eq.(75) as a sequence of the nonselective unitary operations and the
selective phase-shift unitary operations? Choose suitably the algebraic
irrational point $\omega $ and the numbers $M$ and $l$ so that the error $%
O_{0}(M,l)$ in Eq.(75) can be neglected. Then the Trotter-Suzuki theory
[39-41] is exploited to decompose the exponential operator of Eq.(75) into a
sequence of the these unitary operations.

For simplicity, the unitary operation $\exp (-i\theta a_{j}^{s}I_{jz})$ of
Eq.(75) is rewritten as

$\exp (-i\theta a_{j}^{s}I_{jz})=\exp \{-it\stackrel{n-1}{\stackunder{%
k^{\prime }=0}{\sum }}\ \stackrel{M\times l}{\stackunder{k=-M\times l}{\sum }%
}A_{j}(k,k^{\prime },\mathbf{\omega })\}$

$\qquad \qquad \qquad \times \exp \{-it\stackrel{n-1}{\stackunder{k^{\prime
}=0}{\sum }}\ \stackrel{M\times l}{\stackunder{k=-M\times l}{\sum }}%
A_{j}(k,k^{\prime },-\mathbf{\omega })\}+O_{0}(M,l)\qquad \ \ (79)$ \newline
where $t=\theta /2$ due to the fact that $\left\| 2a_{j}^{s}I_{jz}\right\|
=1,$ and the Hermitian operator $A_{j}(k,k^{\prime },\mathbf{\omega })$ is
written as

$A_{j}(k,k^{\prime },\mathbf{\omega })=\dfrac{\Phi (M,l,k)}{8n(2M+1)^{l}}%
\exp (-i\varphi _{jk^{\prime }}F_{jz})$

$\qquad \qquad \times G_{jQ}(k\omega _{1},k\omega _{2},...,k\omega _{n})\exp
(i\varphi _{jk^{\prime }}F_{jz}).\qquad \qquad \qquad \qquad \quad \ (80)$ 
\newline
With the help of the generalized Trotter formula [39] the unitary operation $%
\exp (-i\theta a_{j}^{s}I_{jz})$ of Eq.(79) can be decomposed approximately
as

$\exp (-i\theta a_{j}^{s}I_{jz})=\{\stackrel{n-1}{\stackunder{k^{\prime }=0}{%
\prod }}\ \stackrel{M\times l}{\stackunder{k=-M\times l}{\prod }}\exp [-it\
A_{j}(k,k^{\prime },\mathbf{\omega })/L_{0}]\}^{L_{0}}$

$\times \{\stackrel{n-1}{\stackunder{k^{\prime }=0}{\prod }}\ \stackrel{%
M\times l}{\stackunder{k=-M\times l}{\prod }}\exp [-itA_{j}(k,k^{\prime },-%
\mathbf{\omega })/L_{0}]\}^{L_{0}}+O_{0}(M,l)+O(L_{0})\qquad \ (81)$ \newline
where the error function operator $O(L_{0})$ has the upper bound [11]:

$\qquad \qquad \qquad \left\| L_{0}[\exp (-i\theta
a_{j}^{s}I_{jz}/L_{0})-1+i\theta a_{j}^{s}I_{jz}/L_{0}]\right\| $, \newline
and by Eq.(80) the unitary operator $\exp [-itpA_{j}(k,k^{\prime },\mathbf{%
\omega })]$ with any real constant $p$ can be explicitly expressed as

$\exp [-itpA_{j}(k,k^{\prime },\mathbf{\omega })]=\exp (-i\varphi
_{jk^{\prime }}F_{jz})$

$\times \exp [-itp\dfrac{\Phi (M,l,k)}{8n(2M+1)^{l}}G_{jQ}(k\omega
_{1},k\omega _{2},...,k\omega _{n})]\exp (i\varphi _{jk^{\prime
}}F_{jz}).\qquad \quad \ \ (82)$ \newline
The number of the unitary operations $\exp [-itA_{j}(k,k^{\prime },\mathbf{%
\omega })/L_{0}]$ in Eq.(81) are $2L_{0}n(2M\times l+1).$ It follows from
Eq.(81) that the oracle unitary operation $U_{oz}(\theta )$ is written as

$U_{oz}(\theta )=\stackrel{n}{\stackunder{j=1}{\prod }}\exp [-i\theta
a_{j}^{s}I_{jz}]$

$\qquad \quad =\stackrel{n}{\stackunder{j=1}{\prod }}[\{\stackrel{n-1}{%
\stackunder{k^{\prime }=0}{\prod }}\ \stackrel{M\times l}{\stackunder{%
k=-M\times l}{\prod }}\exp [-it\ A_{j}(k,k^{\prime },\mathbf{\omega }%
)/L_{0}]\}^{L_{0}}$

$\times \{\stackrel{n-1}{\stackunder{k^{\prime }=0}{\prod }}\ \stackrel{%
M\times l}{\stackunder{k=-M\times l}{\prod }}\exp [-itA_{j}(k,k^{\prime },-%
\mathbf{\omega })/L_{0}]\}^{L_{0}}]+nO_{0}(M,l)+nO(L_{0})\qquad (83)$ 
\newline
where the error function operator $nO(L_{0})$ is less than the upper bound
[11]:

\qquad \qquad \qquad $\stackrel{n}{\stackunder{j=1}{\max }}\{\left\|
nL_{0}[\exp (-i\theta a_{j}^{s}I_{jz}/L_{0})-1+i\theta
a_{j}^{s}I_{jz}/L_{0}]\right\| \}$. \newline
This error function can be made as small as desired when $L_{0}$ is taken as
a sufficiently large number, ensuring that the oracle unitary operation $%
U_{oz}(\theta )$ can be expressed correctly as a sequence of the unitary
operations $\exp [-itA_{j}(k,k^{\prime },\mathbf{\omega })$ $/L_{0}]$ as
Eq.(83) within the desired small error (here the error $nO_{0}(M,l)$ can be
neglected, see Eq.(78b)). Note that $U_{oy}(\theta )=\exp (i\frac{\pi }{2}%
F_{x})U_{oz}(\theta )\exp (-i\frac{\pi }{2}F_{x}).$ By inserting the oracle
unitary operation $U_{oz}(\theta )$ ($\theta =\pi /4$) of Eq.(83) into
Eq.(34) one obtains finally the real quantum search network $U_{S}$ (34)\
which is expressed as a sequence of the nonselective unitary operations and
the selective phase-shift unitary operations $C_{s}(\theta )$.

In fact, the decomposition (81) for the unitary operation $\exp (-i\theta
a_{j}^{s}I_{jz})$ is the first-order approximated decomposition. A more
accurate decomposition, that is, the higher order approximated
decomposition, may be achieved with the help of the Suzuki theory [39-41].
As suggested by Suzuki [39-41], the exponential operators on the right-hand
side of Eq.(79) is first decomposed approximately into a symmetric sequence
of simple exponential operators $\exp [-itpA_{j}(k,k^{\prime },\mathbf{%
\omega })]$ $(\left| p\right| <1)$ with the error of order $O(t^{2m})$

$\exp \{-it\stackrel{n-1}{\stackunder{k^{\prime }=0}{\sum }}\ \stackrel{%
M\times l}{\stackunder{k=-M\times l}{\sum }}A_{j}(k,k^{\prime },\mathbf{%
\omega })\}=f_{2m-1}(\{A_{j}(k,k^{\prime },\mathbf{\omega })\})+O(t^{2m}).\
\ (84)$ \newline
The $(2m-1)$-order approximated symmetrized decomposition $f_{2m-1}$ in
Eq.(84) can be constructed as

$\qquad \qquad f_{2m-1}(\{A_{j}(k,k^{\prime },\mathbf{\omega })\})=\stackrel{%
R}{\stackunder{j^{\prime }=1}{\prod }}S(itp_{2m-1j^{\prime }})\qquad \qquad
\qquad \qquad \ \ \ (85)$ \newline
with the 2-order symmetrized decomposition $S(it)$:

$S(it)=\exp [-itA_{j}(-M\times l,0,\mathbf{\omega })/2]......\exp
[-itA_{j}(M\times l,n-2,\mathbf{\omega })/2]$

$\times \exp [-itA_{j}(M\times l,n-1,\mathbf{\omega })]\exp
[-itA_{j}(M\times l,n-2,\mathbf{\omega })/2]......$

$\times \exp [-itA_{j}(-M\times l,0,\mathbf{\omega })/2]\qquad \qquad \qquad
\qquad \qquad \qquad \qquad \qquad \quad \ \ (86)$ \newline
where the parameters $\{p_{2m-1j^{\prime }}\}$ are normalized, $\stackrel{R}{%
\stackunder{j^{\prime }=1}{\sum }}p_{2m-1j^{\prime }}=1$, and it has been
shown that the $2m$-order symmetrized decomposition really equals to the $%
(2m-1)$-order one [39]. The explicit determination for the parameters $%
\{p_{2m-1j^{\prime }}\}$ in Eq.(85) is given in Refs.[39, 40]. By exploiting
the generalized Trotter-Suzuki formula [39, 40] one can obtain more accurate
symmetrized decomposition with the smaller error of order $%
O(L^{-(2m-1)}t^{2m})$ from Eq.(84)

$\exp \{-it\stackrel{n-1}{\stackunder{k^{\prime }=0}{\sum }}\ \stackrel{%
M\times l}{\stackunder{k=-M\times l}{\sum }}A_{j}(k,k^{\prime },\mathbf{%
\omega })\}=[f_{2m-1}(\{\frac{1}{L}A_{j}(k,k^{\prime },\mathbf{\omega }%
)\})]^{L}$

$\qquad \qquad \qquad \qquad +O(t^{2m}/L^{(2m-1)}).\qquad \qquad \qquad
\qquad \qquad \qquad \qquad \ (87)$ \newline
There are $[2n(2M\times l+1)-1]$ unitary operators $\exp
[-itpA_{j}(k,k^{\prime },\mathbf{\omega })]$ for each two-order symmetrized
decomposition $S(it)$ of Eq.(86)$.$ Then there are $R[2n(2M\times l+1)-1]$
unitary operators $\exp [-itpA_{j}(k,k^{\prime },\mathbf{\omega })]$ in the $%
(2m-1)$-order symmetrized decomposition $f_{2m-1}(\{A_{j}(k,k^{\prime },%
\mathbf{\omega })\})$ of Eq.(85) and the symmetrized decomposition of
Eq.(87) contains the unitary operators with the total number of $%
LR[2n(2M\times l+1)-1].$

Now inserting the symmetrized decomposition (87) into Eq.(79) the unitary
operator $\exp (-i\theta a_{j}^{s}I_{jz})$ is expressed as

$\exp (-i\theta a_{j}^{s}I_{jz})=[f_{2m-1}(\{\frac{1}{L}A_{j}(k,k^{\prime },%
\mathbf{\omega })\})]^{L}[f_{2m-1}(\{\frac{1}{L}A_{j}(k,k^{\prime },-\mathbf{%
\omega })\})]^{L}$

$\quad \quad \quad +O_{0}(M,l)+O(t^{2m}/L^{(2m-1)}).\quad \qquad \qquad
\qquad \qquad \qquad \qquad \quad \ \ (88)$ \newline
As the final result, the oracle unitary operation $U_{oz}(\theta )$ can be
expressed as a sequence of the unitary operations $\exp
[-itpA_{j}(k,k^{\prime },\mathbf{\omega })]$ which consist of the selective
phase-shift operatons and the nonselective unitary operations$,$

$U_{oz}(\theta )=\stackrel{n}{\stackunder{j=1}{\prod }}\{[f_{2m-1}(\{\frac{1%
}{L}A_{j}(k,k^{\prime },\mathbf{\omega })\})]^{L}[f_{2m-1}(\{\frac{1}{L}%
A_{j}(k,k^{\prime },-\mathbf{\omega })\})]^{L}\}$

$\qquad \qquad \qquad +nO_{0}(M,l)+nO(t^{2m}/L^{(2m-1)}).\qquad \qquad
\qquad \qquad \qquad (89)$

It has been shown that the decomposition of Eq.(89)\ is much more accurate
than that of Eq.(83) and thus, the number of the exponential unitary
operations $\exp [-itpA_{j}(k,k^{\prime },\mathbf{\omega })]$ in Eq.(89) is
much less than that one in Eq.(83) with the same error in magnitude [39].
Now number of the selective phase-shift operations $C_{s}(\theta )$ to
compose the oracle unitary operation $U_{oz}(\theta )$ can be evaluated from
the expression (89). The oracle unitary operation $U_{oz}(\theta )$ contains 
$2nLR[2n(2M\times l+1)-1]$ unitary operations $\exp [-itpA_{j}(k,k^{\prime },%
\mathbf{\omega })]$, whereas equations (46), (66), and (82) show that each
unitary operation $\exp [-itpA_{j}(k,k^{\prime },\mathbf{\omega })]$
contains four selective phase-shift operations $C_{s}(\theta ).$ Therefore,
the oracle unitary operation $U_{oz}(\theta )$ really consists of the number 
$8nLR[2n(2M\times l+1)-1]$ of the selective phase-shift operations $%
C_{s}(\theta )$ in addition to the nonselective unitary operations. On the
other hand, it can be seen from Eqs.(46), (66), (80), (82)-(89) that all the
nonselective unitary operations including $\exp (\pm i\pi F_{y})$, $\exp
(\pm i\frac{\pi }{2}F_{jy}),$ $\exp (\pm i\varphi _{jk}F_{jz})$, and $\exp
(\pm i2\pi \stackrel{n}{\stackunder{k=1,k\neq j}{\sum }}\varphi _{k}I_{kz}),$
etc., (note that the index $j$ runs over all $n$ particles in the quantum
system so that $\exp (\pm i\frac{\pi }{2}F_{jy})$, etc., become the
nonselective unitary operations) in the oracle unitary operation $%
U_{oz}(\theta )$ are one-qubit unitary operations and the total number of
these nonselective unitary operations is proportional to the number $%
nLR[2n(2M\times l+1)-1]$. Therefore, the oracle unitary operation $%
U_{oz}(\theta )$ can be really expressed explicitly as a sequence of the
number $O(nLR[2n(2M\times l+1)-1])$ of the selective phase-shift operations $%
C_{s}(\theta )$ and the number $O(nLR[2n(2M\times l+1)-1])$ of the
nonselective unitary operations. Obviously, the quantum network $U_{S}$ of
Eq.(34) with the oracle unitary operation (89) or (83) could be a
polynomial-time quantum search network only when the lattice point number $%
(2M\times l+1)$ of the numerical integration (65) does not increase
exponentially as the dimensional size $n$ of the multiple integral (63). 
\newline
\newline
\textbf{8. Discussion}

As shown in previous sections, the multiple-quantum operator algebra
formalism has been exploited to construct explicitly a real quantum search
algorithm. In an unsorted search problem the initial state, i.e., the
superposition, and the final state, that is, the target state, are usually
given and fixed in a quantum system. Then the propagator and its
corresponding effective Hamiltonian can be constructed explicitly that
describe in quantum mechanics the time evolution of the quantum system from
the initial state to the final of the search problem. A real quantum search
algorithm could be built up by starting out such propagator, although the
propagator may usually not be compatible with the mathematical structure and
principle of the search problem and hence is not a real quantum search
network. There are two families of elementary unitary operations, that is,
the nonselective unitary operations and the oracle unitary operations, for
example, the selective phase-shift operations, in the unsorted quantum
search problem. These elementary unitary operations are compatible with the
mathematical structure and principle of the search problem. Then, the
propagator is compatible with the mathematical structure and principle of
the search problem and becomes a real quantum search network when it is
expressed explicitly as a sequence of the nonselective and the selective
unitary operations. The multiple-quantum operator algebra formalism plays an
important role in the general construction of the quantum search algorithm.
In particular, the discrete Fourier analysis and the phase cycling technique
based on the characteristic transformation behavior of multiple-quantum
coherence operators under the z-axis rotations are very helpful for the
construction of the quantum search networks.

An unsorted search problem in a large unsorted database is a hard problem in
classical computation and there has not been any efficient classical search
algorithm to solve the NP problem in polynomial time so far. Grover has
showed that the search problem can be fast solved by a quadratically
speed-up quantum search algorithm over any classical algorithms [14].
However, the Grover algorithm is not really a polynomial-time quantum search
algorithm. In the paper two possible schemes are proposed to solve the
unsorted search problem. One scheme is based on the NMR devices [24] (see
Appendix A and C). The oracle unitary operations $U_{0p}(\theta )$ $%
(p=x,y,z) $ in the quantum network $U_{S}$ of Eq.(34) could be implemented
directly on the NMR devices and hence the quantum network $U_{S}$ becomes a
real unsorted quantum search network. With the help of the NMR device [24]
the unsorted search problem which is an NP-problem is converted into a
special knapsack problem which can be solved efficiently in polynomial time
(see Appendix A), indicating that the unsorted search problem could be
solved efficiently on the NMR quantum computer in polynomial time. Another
is that the oracle unitary operations $U_{0p}(\theta )$ are expressed as a
sequence of the nonselective unitary operations and the selective
phase-shift operations which can be implemented directly on an oracle
universal quantum computer with the help of the phase cycling technique, the
numerical multidimensional integration and the Trotter-Suzuki theory. It has
been shown that the computational complexity of the unsorted quantum search
algorithm is dependent mainly upon that of the numerical multidimensional
integration. The proper numerical multidimensional integration methods
should satisfy the requirement that the lattice point number to numerically
integrate the multidimensional integral (63) does not increase exponentially
as the dimensional size of the multiple integral within the desired error so
that the constructed quantum search algorithm becomes an efficient
algorithm. One of the possible numerical methods of multiple integration
[30-37] to evaluate the multiple integral (63) may be the Hua-Wang
number-theoretic method [31, 32]. It is believed that with the Hua-Wang
method the lattice point number of numerical integration does not increase
exponentially as the dimensional size of the multiple integral (63). As a
consequence, it is believed that the quantum network $U_{S}$ of Eq.(34)
could be an efficient quantum search network on an oracle universal quantum
computer. Therefore, it is believed that quantum computers could solve
efficiently a general NP-problem in polynomial time. \newline
\newline
\textbf{Acknowledgment }

Author thanks Prof. M.Suzuki kindly sent his papers [39-41] to the author. 
\newline
\newline
\textbf{References}\newline
1. P.Benioff, The computer as a physical system: A microscopic quantum
mechanical Hamiltonian model of computers as represented by Turing machines,
J.Statist.Phys. 22, 563 (1980)\newline
2. P.Benioff, Quantum mechanical Hamiltonian models of Turing machines,
J.Statist.Phys. 29, 515 (1982)\newline
3. R.Feynman, Simulating physics with computers, Internat.J.Theoret.Phys.
21, 467 (1982); Quantum mechanical computers, Found.Phys. 16, 507 (1986) 
\newline
4. D.Deutsch, Quantum theory, the Church-Turing principle and the universal
quantum computer, Proc.Roy.Soc.Lond. A 400, 97 (1985) \newline
5. D.Deutsch, Quantum computational networks, Proc.Roy.Soc.Lond. A 425, 73
(1989) \newline
6. D.Deutsch and R.Jozsa, Rapid solution of problems by quantum computation,
Proc.Roy.Soc.Lond. A 439, 553 (1992) \newline
7. A.Berthiaume and G.Brassard, Oracle quantum computing, J.Mod.Opt. 41,
2521 (1994) \newline
8. E.Bernstein and U.Vazirani, Quantum complexity theory, SIAM J. Computing,
26, 1411(1997) \newline
9. D.R.Simon, On the power of quantum computation, SIAM J.Computing, 26,
1474 (1997) \newline
10. P.W.Shor, Polynomial-time algorithms for prime factorization and
discrete logarithms on a quantum computer, SIAM J.Computing.. 26, 1484
(1997) \newline
11. S.Lloyd, Universal quantum simulators, Science 273, 1073 (1996) \newline
12. C.H.Bennett, E.Bernstein, G.Brassard, and U.Vazirani, Strengths and
weaknesses of quantum computing, SIAM J.Computing.. 26, 1510 (1997) \newline
13. L.K.Grover, Beyond factorization and search, Science 281, 792 (1998)%
\newline
14. L.K.Grover, Quantum mechanics helps in searching for a needle in a
haystack, Phys.Rev.Lett. 79, 325 (1997) \newline
15. M.Boyer, G.Brassard, P.Hoyer, and A.Tapp, Tight bounds on quantum
searching, http://xxx.lanl.gov/quant-ph/9605034 (1996)\newline
16. L.K.Grover, A framework for fast quantum mechanical algorithms,\newline
http://xxx.lanl.gov/abs/quant-ph/9711043 (1997)\newline
17. N.J.Cerf, L.K.Grover, and C.P.Williams, Nested quantum search and
structured problems, Phys.Rev. A 61, 032303-1 (2000) \newline
18. X.Miao, Multiple-quantum operator algebra spaces and description for the
unitary time evolution of multilevel spin systems, Molec.Phys. 98, 625
(2000) \newline
19. X.Miao, Universal construction of unitary transformation of quantum
computation with one- and two-body interactions, \newline
http://xxx.lanl.gov/abs/quant-ph/0003068 (2000) \newline
20. (a) X.Miao, Universal construction of quantum computational networks in
superconducting Josephson junctions, \newline
http://xxx.lanl.gov/abs/quant-ph/0003113 (2000)

(b) X.Miao, A convenient method to prepare the effective pure state in a
quantum ensemble, http://xxx.lanl.gov/abs/quant-ph/0008094 (2000) \newline
21. A.Yao, Quantum circuit complexity, Proc. 34th Annual IEEE Symposium on
Foundations of Computer Science, IEEE Press, Piscataway, NJ, 1993, pp.352%
\newline
22. H.Bennett, Logical reversibility of computation, IBM J.Res.Develop. 17,
525 (1973)\newline
23. R.R.Ernst, G.Bodenhausen, and A.Wokaun, Principles of Nuclear Magnetic
Resonance in One and Two Dimensions (Oxford University Press, Oxford, 1987)%
\newline
24. Z.L.Madi, R.Bruschweiler, and R.R.Ernst, One- and two-dimensional
ensemble quantum computing in spin Liouville space, J.Chem.Phys. 109, 10603
(1998) \newline
25. D.Beckman, A.N.Chari, S.Devabhaktuni, and J.Preskill, Efficient networks
for quantum factoring, Phys.Rev. A 54, 1034 (1996) \newline
26. R.Jozsa, Quantum algorithm and the Fourier transform, Proc.Roy.Soc.
Lond. A 454, 323 (1998)\newline
27. G.Bodenhausen, H.Kogler, and R.R.Ernst, J.Magn.Reson. 58, 370 (1984)%
\newline
28. E.O.Brigham, The fast Fourier transform and its applications
(Prentice-Hall, Englewood Cliffs, N.J., 1988) \newline
29. Yue-sheng Li and You-qian Huang, Numerical approximation (in Chinese)
(the People's Education Press, Beijing, 1978)\newline
30. Li-zhi Xu and Yun-shi Zhou, The high-dimensional numerical integrations
(in Chinese) (Science Press, Beijing, 1980) \newline
31. L.K.Hua and Y.Wang, On uniform distribution and numerical analysis
(Numerical-theoretical method). I, II, III, Sci.Sinica 16, 483 (1973); 17,
331 (1974); 18, 184 (1975) \newline
32. Loo Keng Hua and Yuan Wang, Applications of number theory to numerical
analysis (Springer-Verlag, Berlin, 1981); Numerical integration and its
applications (in Chinese) (Science Press, Beijing, 1963) \newline
33. H.Niederreiter, Quasi-Monte Carlo methods and pseudo-random numbers,
Bull.Amer.Math.Soc. 84, 957 (1978) \newline
34. H.Niederreiter, On a number-theoretical integration method, Aequationes
Math. 8, 304 (1972)\newline
35. H.Niederreiter, Application of diophantine approximations to numerical
integration, Diophantine Application and Its Applications (C.F.Osgood, ed.,
Academic Press, New York, 1973, pp.129-199) \newline
36. T.Tsuda, Numerical integration of functions of very many variables,
Numer.Math. 20, 377 (1973) \newline
37. Loo Keng Hua, An introduction to number theory (in Chinese) (Science
Press, Beijing, 1957) \newline
38. W.M.Schmidt, Simultaneous approximation to algebraic numbers by
rationals, Acta Math. 125, 189 (1970) \newline
39. M.Suzuki, Decomposition formulas of exponential operators and Lie
exponentials with some applications to quantum mechanics and statistical
physics, J.Math.Phys. 26, 601 (1985) \newline
40. M.Suzuki, Fractal decomposition of exponential operators with
applications to many-body theories and Monte-Carlo simulations, Phys.Lett. A
146, 319 (1990); General theory of higher-order decomposition of exponential
operators and symplectic integrators, Phys.Lett. A 165, 387 (1992) \newline
41. M.Suzuki, Convergence of general decompositions of exponential
operators, Commun.Math.Phys. 163, 491 (1994) \newline
42. R.C.Merkle and M.E.Hellman, Hiding information and signatures in
trapdoor knapsacks, IEEE Trans.Inform.Theory, IT-24, 525 (1978) \newline
43. E.Horowitz and S.Sahni, Computing partitions with applications to the
Knapsack problem, JACM, 21, 277 (1974) \newline
44. O.H.Ibarra and C.E.Kim, Fast approximation algorithms for the knapsack
and sum of subset problems, JACM, 22, 463 (1975) \newline
45. E.Farhi, J.Goldstone, S.Gutmann, and M.Sipser, A limit on the speed of
quantum computation in determining parity, Phys.Rev.Lett. 81, 5442 (1998)%
\newline
46. N.A.Gershenfeld and I.L.Chuang, Science 275, 350 (1997); D.G.Cory,
A.F.Fahmy, T.F.Havel, Proc.Natl.Acad.Sci. USA 94, 1634 (1997)\newline
47. S.Macura, Y.Huang, D.Suter, and R.R.Ernst, J.Magn.Reson. 43, 259 (1981);
O.W.Sorensen, M.Rance, and R.R.Ernst, J.Magn.Reson. 56, 527 (1984);
A.L.Davis, G.Estcourt, J.Keeler, E.D.Laue, and J.J.Titman, \newline
J.Magn.Reson. A105, 167 (1993)\newline
48. W.S.Warren; N.Gershenfeld and I.L.Chuang, Science 277, 1688-1690 (1997)%
\newline
49. I.L.Chuang, N.A.Gershenfeld, and M.Kubinec, Phys.Rev.Lett. 80, 3408
(1998); J.A.Jones, M.Mosca, R.H.Hansen, Nature 393, 344 (1998)\newline
\newline
\textbf{Appendix A}

An NMR model based on the spectral labeling [24] is proposed to solve
efficiently the unsorted search problem experimentally. It is based on the
massive parallelism of quantum computation and the noncollapse,
nondemolition, and phase-sensitive measurement in NMR techniques. This NMR
model could offer the possibility to transform the hard NP-problem of the
unsorted search problem into the polynomial-time problem, resulting in that
the hard problem could be solved efficiently. This result supports the
belief that a quantum computer could solve in principle a general
NP-problem, although the present model could work only on a few-qubit
system. The spectral labeling on the NMR quantum computing is first proposed
by Madi, Bruschweiler, and Ernst [24]. One of the advantages of the method
over the other state labelings is that each resonance peak of the NMR
spectrum of the ancillary spin corresponds one-to-one to a quantum state of
the work space of quantum computation. This labeling method requires that
the ancillary spin be coupled with all the spins in the work space and all
single-quantum transitions of the ancillary spin be nondegenerate. This may
limit the practical application of the model in an NMR system with many
qubits. However, it provides a very simple experiment model to solve
efficiently the unsorted search problem.

Consider the weakly coupled spin (I=1/2) system SAMX... as a quantum
computation device, that is, an NMR quantum computer. The spin S is the
ancillary qubit and the spins A, M, X, ... form the work space of the
quantum computation. In particular, assume that all the spins A, M, X, ...
couple with the spin S. In order to exploit the massive parallelism of
quantum computation the system is first prepared at a superposition
including the marked state $|s\rangle $:

$\qquad |\Psi \rangle =|r,S\rangle =\stackrel{N-1}{\stackunder{x=0}{\sum }}%
a_{x}|x\rangle [\frac{1}{\sqrt{2}}(|0\rangle -|1\rangle )]$ \newline
where the ancillary qubit $S$ is at the superposition $\frac{1}{\sqrt{2}}%
(|0\rangle -|1\rangle ).$ The evaluation of the function $f(x)$ then can be
achieved by performing the oracle unitary operation $U_{f}$ on the
superposition

$U_{f}:\ \ |\Psi \rangle \rightarrow \stackrel{N-1}{\stackunder{x=0}{\sum }}%
a_{x}|x\rangle [\frac{1}{\sqrt{2}}(|0\bigoplus f(x)\rangle -|1\bigoplus
f(x)\rangle )]$

$\qquad \qquad =\stackrel{N-1}{\stackunder{x=0,x\neq s}{\sum }}%
a_{x}|x\rangle [\frac{1}{\sqrt{2}}(|0\rangle -|1\rangle )]-a_{s}|s\rangle [%
\frac{1}{\sqrt{2}}(|0\rangle -|1\rangle )]$ \newline
where it has been introduced the fact that the function $f(s)=1$ and $f(r)=0$%
, $r\neq s$ . Obviously, only the target state $|s\rangle [\frac{1}{\sqrt{2}}%
(|0\rangle -|1\rangle )]$ is inverted and any other states keep unchanged
when performing each evaluation of the function $f(x).$ In general, it is
difficult to distinguish the inverted target state from the other states in
a quantum system after performing only once evaluation of the function $f(x)$%
. This is because each state has the equal probability. However, there is a
significant phase difference between the target state and any other state
after implementing once evaluation of the function $f(x)$. The phase of the
target state is opposite to all of the other states. In the Grover quantum
search algorithm [14] this phase difference is transferred into an amplitude
difference between the target state and the other states by making the
diffusion transformation so that the amplitude of the target state is
amplified, resulting in that the search is quadratically speeded up. It is
well-known that the phase difference among states in a nuclear spin ensemble
may be detected by the nuclear magnetic resonance (NMR) measurement
techniques. The NMR signals and spectra carry the information of the phase
difference. Now, it is possible to reveal the effect of the phase difference
on the NMR spin ensemble, and especially on the phase and amplitude of the
NMR spectra if the evaluation of the function $f(x)$ is performed on an NMR
quantum computer. This phase difference may result in the phase-inversion
spectrum of the target state with respect to those peaks of the other
states. As an example, Figure One (see central spectrum) is the conventional
NMR spectrum of the ancillary spin $S$ in a four-qubit spin system SAMX.
When the evaluation of the function $f(s)$ ($|s\rangle =|\alpha \beta \beta
\rangle $) is carried out the target state $|\alpha \beta \beta \rangle
|S\rangle $ is inverted in phase. Now, if the NMR spectrum of the ancillary
qubit $S$ is recorded one may find the phase-inversion peak of the target
state (see bottom spectrum). It shows how the resonant peak of the target
state with inversion phase may be recognized from the others peaks. The
resonant frequency of the phase-inversion peak can be measured accurately if
the signal-to-noise ratio of the peak is high enough. The resonant frequency
actually carries some information of the target state. It is expected to
extract the information from the resonant frequency.

The spin Hamiltonian of the weakly coupled spin (I=1/2) system SAMX... can
be written generally as

$H=\Omega _{S}S_{z}+\stackrel{n}{\stackunder{k=1}{\sum }}\Omega
_{k}I_{kz}+S_{z}\stackrel{n}{\stackunder{k=1}{\sum }}J_{Sk}I_{kz}+%
\stackunder{l>k=1}{\stackrel{n}{\sum }}J_{kl}I_{kz}I_{lz}\qquad \qquad
\qquad \ (A1)$ \newline
where the contribution of the decoherence has been neglected; the symbol I
denotes the spins A, M, X, ..., and $\Omega _{S},$ $\Omega _{k}$ are the
chemical shifts of the spin $S$ and the $k$th spin I, respectively; $J_{Sk}$
is the scalar coupling constant between the spin S and the $k$th spin I, $%
J_{kl}$ the scalar coupling constant of the $k$th and the $l$th spin I.
Obviously, the conventional quantum computational base $|r,S\rangle
=|r\rangle |S\rangle $ are the eigenbase of the spin Hamiltonian (A1) with
the corresponding energy eigenvalue $E(r,S)$,

$\qquad \qquad \qquad \qquad H|r\rangle |S\rangle =E(r,S)|r\rangle |S\rangle
,$

$E(r,S)=\frac{1}{2}b_{S}(\Omega _{S}+\stackrel{n}{\stackunder{k=1}{\sum }}%
\frac{1}{2}a_{k}^{r}J_{Sk})+\stackrel{n}{\stackunder{k=1}{\sum }}\frac{1}{2}%
a_{k}^{r}\Omega _{k}+\stackrel{n}{\stackunder{l>k=1}{\sum }}\frac{1}{4}%
a_{k}^{r}a_{l}^{r}J_{kl}\qquad \quad \ \ (A2)$ \newline
where the unity-number representation of the eigenbase $|r,S\rangle $ has
been used (see Eq.(18)),

$|r,S\rangle =(\frac{1}{2}T_{1}+a_{1}^{r}S_{1})\bigotimes (\frac{1}{2}%
T_{2}+a_{2}^{r}S_{2})\bigotimes ...\bigotimes (\frac{1}{2}%
T_{n}+a_{n}^{r}S_{n})\bigotimes (\frac{1}{2}T_{S}+b_{S}S_{S}).$\newline
For an arbitrary computational basis $|r\rangle $ of spins $I$ the
transition frequency of the ancillary spin $S$ is written as

$\omega _{S}(r)=E(r,S=+1/2)-E(r,S=-1/2)=\Omega _{S}+\stackrel{n}{\stackunder{%
k=1}{\sum }}\frac{1}{2}a_{k}^{r}J_{Sk}.\qquad \quad (A3)$ \newline
The scalar coupling constants $J_{Sk}$ and the chemical shift $\Omega _{S}$
are usually fixed for a given weakly coupled spin system. If one can measure
exactly the resonant frequency $\omega _{S}(r)$ in an NMR experiment the
unity-number vector $\{a_{k}^{r}\}$ will be determined from the above
equation (A3) which can be reduced to the form

$\qquad \qquad \qquad \qquad \stackrel{n}{\stackunder{k=1}{\sum }}%
b_{k}^{r}J_{Sk}=$ $f_{S}(r)\qquad \qquad \qquad \qquad \qquad \qquad \qquad
(A4)$\newline
where $b_{k}^{r}=\frac{1}{2}(a_{k}^{r}+1)=0,1;$ $k=1,2,..,n$ and $%
f_{S}(r)=\omega _{S}(r)-\Omega _{S}+\stackrel{n}{\stackunder{k=1}{\sum }}%
\frac{1}{2}J_{Sk}.$ Obviously, it is the famous knapsack problem to solve
exactly equation (A4). It is well known that the knapsack problem is
generally an NP-complete problem [42, 43]. Then it is usually hard to solve
equation (A4). The degree of difficulty of the problem (A4) is crucially
dependent upon the choice of the coefficients of Eq.(A4), i.e., the scalar
coupling constants $\{J_{Sk}\}$. For example, if the scalar coupling
constant set $\{J_{Sk},k=1,2,...,n\}$ is a superascend sequence, that is, $%
J_{Sk}>0$ $(k=1,2,...,n)$ and

$\qquad \qquad J_{Sk+1}>\stackrel{k}{\stackunder{l=1}{\sum }}J_{Sl},\ (1\leq
k<n-1)$, \qquad \qquad \qquad \newline
equation (A4) can be solved efficiently in polynomial time [42, 43]. Once
the unity number vector $\{a_{k}^{s}\}$ is determined, the oracle unitary
operation $U_{op}(\theta )=\stackrel{n}{\stackunder{k=1}{\prod }}\exp
[-i\theta a_{k}^{s}I_{kp}]$ can be directly implemented experimentally on a
universal quantum computer. Then the target state $|s\rangle $ can be
obtained by making directly the quantum search network $U_{S}$ of Eq.(34) on
the superposition $|\Psi \rangle $ in Eq.(19). It must be pointed out that
there are a number of choices of the coefficients in Eq.(A4) besides the
above superascend sequence so that equation (A4) can be solved efficiently
in polynomial time [42-44].

The above NMR model [24] transforms really the NP-problem, i.e., the
unsorted quantum search problem, to the polynomial-time problem, i.e., the
special knapsack problem [42, 43].\newline
\newline
\textbf{Caption of Figure 1.} A simple NMR device with the weakly coupled
four-spin (I=1/2) system SAMX to solve efficiently the unsorted search
problem. The spin $S$ acts as the auxiliary qubit. Note that the peak of the
target state $|\alpha \beta \beta \rangle |S\rangle $ is inverted in phase
with respect to other peaks, as can be seen in bottom spectrum.\newline
\newline
\newline
\textbf{Appendix B}

If the multidimensional function $f(x_{1},x_{2},...,x_{n})$ can be expanded
as the absolutely convergent Fourier series:

$f(x_{1},x_{2},..,x_{n})=\stackrel{\infty }{\stackunder{m_{1},...m_{n}=-%
\infty }{\sum }}C(m_{1},m_{2},...m_{n})$

$\qquad \qquad \qquad \times \exp [-i2\pi
(m_{1}x_{1}+m_{2}x_{2}+...+m_{n}x_{n})]\ \ \qquad \qquad \qquad (B1)$ 
\newline
with the complex expansion coefficients $C(m_{1},m_{2},...m_{n})$ and
integer indexes $\{m_{k};k=1,2,...,n\},$ and if the integer weight functions 
$\Phi (M,l,j)$ are defined by the following identity

$\qquad \qquad \qquad \quad (\stackrel{M}{\stackunder{j=-M}{\sum }}%
Z^{j})^{l}=\stackrel{M\times l}{\stackunder{j=-M\times l}{\sum }}\Phi
(M,l,j)Z^{j},\qquad \qquad \qquad \quad \ \ (B2)$ \newline
then one has the numerical multiple integration formula:

$\stackrel{1}{\stackunder{0}{\int }}...\stackrel{1}{\stackunder{0}{\int }}%
dx_{1}...dx_{n}f(x_{1},x_{2},..,x_{n})$

$\qquad =\dfrac{1}{(2M+1)^{l}}\stackrel{M\times l}{\stackunder{j=-M\times l}{%
\sum }}\Phi (M,l,j)f(j\mathbf{\omega })+O(M,l)$ \qquad \qquad \qquad \quad $%
(B3)\newline
$with the $n$-dimensional real algebraic number lattice point $\mathbf{%
\omega }=(\omega _{1},\omega _{2},..,\omega _{n}),$ and the error function
operator $O(M,l)$ can be expressed as\newline
$O(M,l)=-\stackrel{\infty }{\stackunder{m_{1},...m_{n}=-\infty }{%
\sum^{\prime }}}C(m_{1},m_{2},...m_{n})(\dfrac{\sin [(2M+1)\pi (\mathbf{m},%
\mathbf{\omega })]}{(2M+1)\sin [\pi (\mathbf{m},\mathbf{\omega })]}%
)^{l}\qquad \ (B4)$ \newline
where the sums $\sum^{\prime }$ with the prime symbol do not include the
term with $m_{1}=m_{2}=...=m_{n}=0$ and the dot product $(\mathbf{m},\mathbf{%
\omega })=m_{1}\omega _{1}+m_{2}\omega _{2}+...+m_{n}\omega _{n}.$\newline
\newline
\textbf{Proof:} The detailed proof for the above theorem can be seen in
references [31, 32]. Note that the multiple integral on the left-hand side
of Eq.(B3) equals the coefficient $C(m_{1},m_{2},...m_{n})$ with $%
m_{1}=m_{2}=...=m_{n}=0,$

$\qquad \qquad C(0,0,...0)=\stackrel{1}{\stackunder{0}{\int }}...\stackrel{1%
}{\stackunder{0}{\int }}dx_{1}...dx_{n}f(x_{1},x_{2},..,x_{n}).\qquad \qquad
\ \qquad \ $ \newline
Because the Fourier series of Eq.(B1) is absolutely convergent one has%
\newline
$\dfrac{1}{(2M+1)^{l}}\stackrel{M\times l}{\stackunder{j=-M\times l}{\sum }}%
\Phi (M,l,j)f(j\mathbf{\omega })=\dfrac{1}{(2M+1)^{l}}\stackrel{\infty }{%
\stackunder{m_{1},...m_{n}=-\infty }{\sum }}C(m_{1},m_{2},...m_{n})$

$\qquad \qquad \qquad \times \stackrel{M\times l}{\stackunder{j=-M\times l}{%
\sum }}\Phi (M,l,j)\exp [-i2\pi j(m_{1}\omega _{1}+m_{2}\omega
_{2}+...+m_{n}\omega _{n})]\ $

$=C(0,0,...,0)+\dfrac{1}{(2M+1)^{l}}\stackrel{\infty }{\stackunder{%
m_{1},...m_{n}=-\infty }{\sum^{\prime }}}C(m_{1},m_{2},...m_{n})$

$\times (\stackrel{M}{\stackunder{j=-M}{\sum }}\exp [-i2\pi j(m_{1}\omega
_{1}+m_{2}\omega _{2}+...+m_{n}\omega _{n})])^{l}$

$=\stackrel{1}{\stackunder{0}{\int }}...\stackrel{1}{\stackunder{0}{\int }}%
dx_{1}...dx_{n}f(x_{1},x_{2},..,x_{n})$

$+\stackrel{\infty }{\stackunder{m_{1},...m_{n}=-\infty }{\sum^{\prime }}}%
C(m_{1},m_{2},...m_{n})(\dfrac{\sin [(2M+1)\pi (\mathbf{m},\mathbf{\omega })]%
}{(2M+1)\sin [\pi (\mathbf{m},\mathbf{\omega })]})^{l}$ $\newline
$where the identity (B2) has been introduced. Obviously, if the error
function $O(M,l)$ is given by Eq.(B4) then the numerical integration formula
(B3) is obtained.

To calculate explicitly the error function operator $O(M,l)$ in the
numerical integration (65) the integrand operator (66) of the multiple
integral (63) is first expanded under any pair of the usual computational
bases $|r\rangle $ and $|t\rangle $ and the corresponding matrix element
then is written as

$\langle r|G_{jq}(y_{1},y_{2},...,y_{n})|t\rangle =\langle r|\theta
H_{jq}|t\rangle \exp [-i2\pi \stackrel{n}{\stackunder{k=1,k\neq j}{\sum }}%
m_{k}(r,t)y_{k}]\qquad \ \ (B5)$ \newline
where the integer vector $\mathbf{m}=\{m_{k}(r,t)\}$ with $%
m_{k}(r,t)=-1,0,+1;$ $k=1,2,...,n$ for any pair of bases $|r\rangle $ and $%
|t\rangle ,$ and $m_{1}(r,t)=m_{2}(r,t)=...=m_{n}(r,t)=0$ when $|r\rangle
=|t\rangle .$ Then, the error function operator $O(M,l)$ in the numerical
integration (65) with the integrand (B5) can be found from Eq.(B4),

$\langle r|O(M,l)|t\rangle =$ $-\langle r|\theta H_{jq}|t\rangle (\dfrac{%
\sin [(2M+1)\pi (\mathbf{m},\mathbf{\omega })]}{(2M+1)\sin [\pi (\mathbf{m},%
\mathbf{\omega })]})^{l}\quad (r\neq t).$ \newline
Obviously, $\langle r|O(M,l)|t\rangle =0$ for any pair of bases $|r\rangle
=|t\rangle .$ \newline
\newline
\newline
\textbf{Appendix C}

In appendix A the NMR device [24] is used to measure experimentally the
unity-number vector $\{a_{k}^{s}\}$ in polynomial time. However, this device
is quite limited in practice. Therefore, one hopes naturally that there is a
convenient NMR device to measure efficiently the unity-number vector. It had
better start at the thermal equilibrium state of an NMR quantum ensemble
instead of the effective pure state [46]. That is, this NMR device can
exploit the initial mix state of a spin system, for example, the thermal
equilibrium state to determine experimentally the unity-number vector and
hence it may be useful in practice. The present device is still based on the
massive parallelism of quantum computation and the phase-sensitive NMR
measurement.

In section 4 it has been shown that the selective phase-shift operation $%
C_{s}(\theta )$ can be equivalent to the oracle unitary operation $U_{f}$
when the auxiliary qubits are used in the implementation of the quantum
search problem. For example, $C_{s}(\pi )=U_{f}$ (equivalent ) when the
auxiliary qubit $S$ takes the superposition $|S\rangle =\QDABOVE{1pt}{1}{%
\sqrt{2}}(|0\rangle -|1\rangle )$ in the superposition:

$\qquad \qquad \qquad |\Psi _{1}\rangle =|I,S\rangle =\ \stackrel{N-1}{%
\stackunder{x=0}{\sum }}a_{x}|x\rangle [\frac{1}{\sqrt{2}}(|0\rangle
-|1\rangle )]\qquad \qquad \qquad \ \ (C1)$ \newline
where the selective phase-shift operation $C_{s}(\pi )$ is applied only to
the work qubits $I$, while the oracle unitary operation $U_{f}$, that is,
the evaluation of function $f(s)$, is applied to all the qubits including
both the work qubits $I$ and the auxiliary qubit $S$. It is also shown that
the selective phase-shift operation $C_{s}(\theta )$ can be expressed as $%
C_{s}(\theta )=U_{f}V(\theta )U_{f}$ (equivalent) on the superposition:

$\qquad \qquad \qquad |\Psi _{2}\rangle =|I,S\rangle =\ \stackrel{N-1}{%
\stackunder{x=0}{\sum }}a_{x}|x\rangle |0\rangle |1\rangle \qquad \qquad
\qquad \qquad \qquad \ \ (C2)$ \newline
where the state of the auxiliary qubits $S$ takes $|S\rangle =|0\rangle
|1\rangle .$ On the other hand, in the matrix representation the
superposition (C1) and (C2) can be expressed respectively as

$\qquad \qquad \qquad \rho _{1}=|\Psi _{1}\rangle \langle \Psi _{1}|=\rho
_{1I}\bigotimes \rho _{1S}\qquad \quad \qquad \qquad \qquad \qquad \quad \
(C3)$ \newline
with $\rho _{1I}=(\stackrel{N-1}{\stackunder{x,y=0}{\sum }}%
a_{x}a_{y}|x\rangle \langle y|)$ and $\rho _{1S}=\QDABOVE{1pt}{1}{2}%
(|0\rangle -|1\rangle )(-\langle 1|+\langle 0|),$ and

$\qquad \qquad \qquad \rho _{2}=|\Psi _{2}\rangle \langle \Psi _{2}|=\rho
_{2I}\bigotimes \rho _{2S}\qquad \qquad \qquad \qquad \qquad \qquad \ (C4)$ 
\newline
with $\rho _{2I}=(\stackrel{N-1}{\stackunder{x,y=0}{\sum }}%
a_{x}a_{y}|x\rangle \langle y|)$ and $\rho _{2S}=|0\rangle \langle
0|\bigotimes |1\rangle \langle 1|.$ Then there are the following relations
when the oracle unitary operations $U_{f}$ and $U_{f}V(\theta )U_{f}$
applied to the two superpositions (C3) and (C4), respectively,

$\qquad \qquad U_{f}\rho _{1}U_{f}^{-1}=C_{s}(\pi )\rho _{1I}C_{s}(\pi
)^{-1}\bigotimes \rho _{1S},$\qquad $\qquad \qquad \qquad \qquad \ (C5)$

$\qquad U_{f}V(\theta )U_{f}\rho _{2}(U_{f}V(\theta
)U_{f})^{-1}=C_{s}(\theta )\rho _{2I}C_{s}(\theta )^{-1}\bigotimes \rho
_{2S} $\qquad $\qquad \quad \ (C6)$ \newline
This shows that the action of the selective phase-shift operations $%
C_{s}(\theta )$ and $C_{s}(\pi )$ on the work qubits $I$ is equivalent to
the action of the oracle unitary operations $U_{f}V(\theta )U_{f}$ and $%
U_{f} $ on the whole system including the auxiliary qubits $S,$
respectively. Although the density operators $\rho _{1}$ and $\rho _{2}$ are
pure states in Eqs.(C5) and (C6), the two equations (C5) and (C6) still hold
even when the density operators $\rho _{1I}$ and $\rho _{2I}$ take any mix
states of the work qubits $I.$ This is because the unitary operations $%
U_{f}, $ $V(\theta ),$ and $C_{s}(\theta )$ are linear operators. Then the
equivalent relations: $C_{s}(\pi )=U_{f}$ and $C_{s}(\theta )=U_{f}V(\theta
)U_{f}$ still hold even when the density operators $\rho _{1I}$ and $\rho
_{2I}$ are taken as the mix states of the work qubits $I$ in Eqs.(C5) and
(C6). This point is important for the NMR experimental determination for the
unity-number vector $\{a_{k}^{s}\}$ by startng out the mix state $\rho _{1I}$
or $\rho _{2I}$ of the work qubits $I$, e.g., the thermal equilibrium state,
instead of the effective pure state.

In NMR experiments to determine the unity-number vector $\{a_{k}^{s}\}$ the
density operators $\rho _{1I}$ and $\rho _{2I}$ may be prepared as any mix
states of the work qubits $I,$ but the auxiliary qubits $S$ must be prepared
as the effective pure states $\rho _{1S}$ and $\rho _{2S},$ respectively. On
the other hand, if the density operators $\rho _{1I}$ and $\rho _{2I}$ do
not include the marked state $|s\rangle $ one has

$\qquad \qquad C_{s}(\pi )\rho _{1I}C_{s}(\pi )^{-1}=\rho _{1I},$\qquad $%
C_{s}(\theta )\rho _{2I}C_{s}(\theta )^{-1}=\rho _{2I}.$\newline
In this case it is impossible to determine experimentally the unity-number
vector $\{a_{k}^{s}\}.$ Thus, it is required that the density operators $%
\rho _{1I}$ and $\rho _{2I}$ be taken as any mix state that include any
given marked state $|s\rangle $. As a simple example, the density operator ($%
\rho _{1I}$ or $\rho _{2I}$) including any given marked state $|s\rangle $
may be taken as

$\qquad \qquad \qquad \rho _{I}=\alpha _{0}E+\stackrel{n}{\stackunder{k=1}{%
\sum }}\varepsilon _{k}I_{k\mu }\qquad (\mu =x,y)\qquad \qquad \qquad \qquad
\ \ (C7)$ \newline
This density operator can be generated from the thermal equilibrium state of
the work qubits: $\rho _{0I}=\alpha _{0}E+\stackrel{n}{\stackunder{k=1}{\sum 
}}\varepsilon _{k}I_{kz}$ by applying a ninety degree pulse.

It is assumed in the following discussion that the auxiliary qubits $S$ are
always prepared as the effective pure states $\rho _{1S}$ and $\rho _{2S},$
respectively, during the action of the oracle unitary operations $U_{f}$ and 
$U_{f}V(\theta )U_{f}$ so that the two oracle unitary operations can be
replaced by their corresponding selective phase-shift operations $C_{s}(\pi
) $ and $C_{s}(\theta ),$ respectively, to describe the evolution process
during the oracle unitary operation

$\rho _{1S}=(\QDABOVE{1pt}{1}{2}|0\rangle \langle 0|+\QDABOVE{1pt}{1}{2}%
|1\rangle \langle 1|-\QDABOVE{1pt}{1}{2}|0\rangle \langle 1|-\QDABOVE{1pt}{1%
}{2}|1\rangle \langle 0|)=\QDABOVE{1pt}{1}{2}E-S_{x},\qquad \qquad \ \ (C8)$

$\rho _{2S}=|0\rangle \langle 0|\bigotimes |1\rangle \langle 1|=(\QDABOVE{1pt%
}{1}{2}E_{1}+S_{1z})\bigotimes (\QDABOVE{1pt}{1}{2}E_{2}-S_{2z})$

$\qquad =\QDABOVE{1pt}{1}{4}E+\QDABOVE{1pt}{1}{2}%
(S_{1z}-S_{2z})-S_{1z}S_{2z}\qquad \qquad \qquad \qquad \qquad \qquad \qquad
\ \ (C9)$ \newline
In order to analyse the evolution process during the oracle unitary
operation the selective phase-shift operation $C_{s}(\theta )$ is decomposed
into a sequence of elementary propagators built up with the base operators
of the $LOMSO$ operator subspace [18,19]

$C_{s}(\theta )=\exp (-i\theta _{s}^{*})\exp [-i\theta _{s}^{*}(\stackrel{n}{%
\stackunder{k=1}{\sum }}a_{k}^{s}2I_{kz})]\exp [-i\theta _{s}^{*}(\stackrel{n%
}{\stackunder{l>k=1}{\sum }}a_{k}^{s}a_{l}^{s}4I_{kz}I_{lz})]$

$\times \exp [-i\theta _{s}^{*}(\stackrel{n}{\stackunder{l>k=1}{\sum }}%
a_{k}^{s}a_{l}^{s}a_{m}^{s}8I_{kz}I_{lz}I_{mz})]....$ \qquad $\qquad \qquad
\qquad \qquad \qquad (C10)$\newline
where $\theta _{s}^{*}=\theta _{s}/N.$ Then it is easy to prove that there
are the basic unitary transformation relations for the oracle unitary
operation

$C_{s}(\theta )I_{kx}C_{s}(\theta )^{-1}=I_{kx}\cos [\theta _{s}^{*}.2(E+%
\stackrel{n}{\stackunder{l=1}{\sum^{\prime }}}a_{l}^{s}2I_{lz}+\stackrel{n}{%
\stackunder{m>l=1}{\sum^{\prime }}}a_{l}^{s}a_{m}^{s}4I_{lz}I_{mz}+...)]$

$\qquad +a_{k}^{s}I_{ky}\sin [\theta _{s}^{*}.2(E+\stackrel{n}{\stackunder{%
l=1}{\sum^{\prime }}}a_{l}^{s}2I_{lz}+\stackrel{n}{\stackunder{m>l=1}{%
\sum^{\prime }}}a_{l}^{s}a_{m}^{s}4I_{lz}I_{mz}+...)]\qquad \ \ (C11a)$

$C_{s}(\theta )I_{ky}C_{s}(\theta )^{-1}=I_{ky}\cos [\theta _{s}^{*}.2(E+%
\stackrel{n}{\stackunder{l=1}{\sum^{\prime }}}a_{l}^{s}2I_{lz}+\stackrel{n}{%
\stackunder{m>l=1}{\sum^{\prime }}}a_{l}^{s}a_{m}^{s}4I_{lz}I_{mz}+...)]$

$\qquad -a_{k}^{s}I_{kx}\sin [\theta _{s}^{*}.2(E+\stackrel{n}{\stackunder{%
l=1}{\sum^{\prime }}}a_{l}^{s}2I_{lz}+\stackrel{n}{\stackunder{m>l=1}{%
\sum^{\prime }}}a_{l}^{s}a_{m}^{s}4I_{lz}I_{mz}+...)]\qquad \ \ (C11b)$ 
\newline
where the sums $\sum^{\prime }$ run over all indexes except the index $k.$
To simplify further the unitary transformation relations (C11a) and (C11b)
the triangular functions of the $LOMSO$\ operator subspace are expanded in
terms of the $LOMSO$ base operators [18]

$\cos [\theta _{s}^{*}.2(E+\stackrel{n}{\stackunder{l=1}{\sum^{\prime }}}%
a_{l}^{s}2I_{lz}+\stackrel{n}{\stackunder{m>l=1}{\sum^{\prime }}}%
a_{l}^{s}a_{m}^{s}4I_{lz}I_{mz}+...)]=\alpha _{0k}^{^{\prime }}E+\stackrel{n%
}{\stackunder{l=1}{\sum^{\prime }}}\Omega _{kl}^{^{\prime }}I_{lz}$

$\qquad +\stackrel{n}{\stackunder{m>l=1}{\sum^{\prime }}}J_{klm}^{^{\prime
}}2I_{lz}I_{mz}+\stackrel{n}{\stackunder{p>m>l=1}{\sum^{\prime }}}%
J_{klmp}^{^{\prime }}4I_{lz}I_{mz}I_{pz}+...\qquad \qquad \quad \ \ (C12a)$

$\sin [\theta _{s}^{*}.2(E+\stackrel{n}{\stackunder{l=1}{\sum^{\prime }}}%
a_{l}^{s}2I_{lz}+\stackrel{n}{\stackunder{m>l=1}{\sum^{\prime }}}%
a_{l}^{s}a_{m}^{s}4I_{lz}I_{mz}+...)]=\alpha _{0k}^{^{\prime \prime }}E+%
\stackrel{n}{\stackunder{l=1}{\sum^{\prime }}}\Omega _{kl}^{^{^{\prime
\prime }}}I_{lz}$

$\qquad +\stackrel{n}{\stackunder{m>l=1}{\sum^{\prime }}}J_{klm}^{^{^{\prime
\prime }}}2I_{lz}I_{mz}+\stackrel{n}{\stackunder{p>m>l=1}{\sum^{\prime }}}%
J_{klmp}^{^{^{\prime \prime }}}4I_{lz}I_{mz}I_{pz}+...\qquad \qquad \quad \
(C12b)$ \newline
Now start the initial state $\rho _{I}$ (C7) ($\mu =x$). By applying the
oracle unitary operation and then a hard $90_{x}^{\circ }$ pulse to the
spins $I$ the state is transferred into

$\rho _{f}=R(90_{x})C_{s}(\theta )\rho _{I}C_{s}(\theta
)^{-1}R(90_{x})^{-1}\bigotimes \rho _{2S}$

$=\alpha _{0}E\bigotimes \rho _{2S}+\stackrel{n}{\stackunder{k=1}{\sum }}%
\varepsilon _{k}I_{kx}\{\alpha _{0k}^{^{\prime }}+\stackrel{n}{\stackunder{%
l=1}{\sum^{\prime }}}\Omega _{kl}^{^{\prime }}(-I_{ly})$

$+\stackrel{n}{\stackunder{m>l=1}{\sum^{\prime }}}J_{klm}^{^{\prime
}}2I_{ly}I_{my}+\stackrel{n}{\stackunder{p>m>l=1}{\sum^{\prime }}}%
J_{klmp}^{^{\prime }}4(-I_{ly}I_{my}I_{py})+...\}\bigotimes \rho _{2S}$

$+\stackrel{n}{\stackunder{k=1}{\sum }}\varepsilon
_{k}a_{k}^{s}I_{kz}\{\alpha _{0k}^{^{\prime \prime }}E+\stackrel{n}{%
\stackunder{l=1}{\sum^{\prime }}}\Omega _{kl}^{^{^{\prime \prime
}}}(-I_{ly})+\stackrel{n}{\stackunder{m>l=1}{\sum^{\prime }}}%
J_{klm}^{^{^{\prime \prime }}}2I_{ly}I_{my}$

$+\stackrel{n}{\stackunder{p>m>l=1}{\sum^{\prime }}}J_{klmp}^{^{^{\prime
\prime }}}4(-I_{ly}I_{my}I_{py})+...\}\bigotimes \rho _{2S}\ \quad \qquad
\qquad \qquad \qquad \quad \ \ \ (C13)$ \newline
This state is quite inconvenient for the NMR maesurement of the unity-number
vector $\{a_{k}^{s}\}$. However, by applying a purge pulse unit [47] such as
z-direction gradient field, z-filter, etc., on the density operator (C13) to
cancel all the multiple-quantum coherences including the zero-quantum
coherences but leave only the longitudinal magnetization and spin order
components unchanged, the density operator (C13) is reduced to the form

$\qquad \rho _{f}=\{\alpha _{0}E+\stackrel{n}{\stackunder{k=1}{\sum }}\alpha
_{0k}^{^{\prime \prime }}\varepsilon _{k}a_{k}^{s}I_{kz}\}\bigotimes \rho
_{2S}\qquad \ \ \qquad \qquad \qquad \qquad \quad \ \ (C14)$ \newline
The parameters $\{\alpha _{0k}^{^{\prime \prime }}\}$ can be obtained
explicitly from Eq.(C12b),\newline
$\alpha _{0k}^{^{\prime \prime }}=\QDABOVE{1pt}{2}{N}\stackunder{%
m_{i},m_{p},m_{q},...}{\sum^{\prime }}\sin \{\QDABOVE{1pt}{2}{N}\theta
_{s}[1+\stackrel{n}{\stackunder{i=1}{\sum^{\prime }}}a_{i}^{s}2m_{i}+%
\stackrel{n}{\stackunder{p>i=1}{\sum^{\prime }}}%
a_{i}^{s}a_{p}^{s}4m_{i}m_{p}+...]\}\quad \ \ (C15)$ \newline
where $m_{i},$ $m_{p},$ $m_{q},...=1/2,$ $-1/2$ and the sums $\sum^{\prime }$
run over all indexes except the index $m_{k}$. It turns out that the
parameters $\alpha _{0k}^{^{\prime \prime }}=1/2^{n-1}$ when the phase angle 
$\theta _{s}=\pi /2.$ Now the density operator $\rho _{f}$ of Eq.(C14) is
quite simple. The unity-number vector $\{a_{k}^{s}\}$ can be measured simply
from the state of Eq.(C14). The measurement may be carried out conveniently
by applying a hard $90_{y}^{\circ }$ pulse to the spins $I$ of the density
operator of Eq.(C14) and then recording the NMR signal of the spins $I$
during decoupling the auxiliary spins $S$. Obviously, the unity-number
vector $\{a_{k}^{s}\}$ may be determined conveniently by recording and
comparing both two NMR spectra of the density operators (C7) and (C14),
respectively.

With respect to the NMR\ device in Appendix A the present NMR device can be
even a spin ensemble of linear molecules with neighbor interaction and
moreover, the auxiliary spins in the device are not required to interact
with all spins of the work qubits. Therefore, the present NMR device is very
convenient one to determine the unity-number vector and may be more useful
in practice. However, the NMR singal of the spins $I$ of the density
operator (C14) is proportional to $1/2^{n-1},$ indicating that it decreases
exponentially as the I-spin number $n$. Thus, the main drawback of this
device is that the NMR signal to determine sufficiently the unity-number
vector $\{a_{k}^{s}\}$ decreases exponentially as the qubit number of the
search problem. This is similar to the NMR quantum computing based on the
effective pure state [48].

Although the NMR signal of the density operator (C14) decreases
exponentially as the qubit number $n$ like the NMR quantum computing on the
effective pure state, it must be pointed out that they have a significant
difference. The present device can start at the thermal equilibrium state of
spin ensemble, while the latter starts at the effective pure state [46, 49].
The present scheme to measure the unity-number vector is polynormial-time
within the realizable size of NMR quantum computation, but the Grover
algorithm based on the pure quantum state [14] or the effective pure state
[49] is a quadratically speed-up method even within the realizable size of
NMR technique. Although in the present NMR device the final NMR signal of
Eq.(C14) may not be detected due to too low signal-to-noise ratio in a spin
system with many qubits, this is the NMR \textit{technique limit} instead of
the\textit{\ principle limit}. Therefore, the final result $\{a_{k}^{s}\}$
will be obtained certainly in polynormial time if the sensitivity of the NMR
signal of Eq.(C14) is improved sufficiently in NMR technique [46]. However,
in the version of the Grover algorithm based on the pure quantum state [14]
or the effective pure state [49] each quantum state in the superposition has
the same probability, then the marked state in the superposition is
impossible to be found certainly after performing once the oracle unitary
operation, i.e., the selective phase-shift operation, due to the limit of
the quantum measurement principle [12] even when the measured signal
sensitivity is increased sufficiently. One method to find certainly the
marked state is to amplify the probability of the marked state but suppress
all the others. This is just the spirit of the Grover algorithm [14].

There is an interesting thing. Suppose that the NMR signal of the density
operator (C14) could be enhanced by applying a sequence of a polynomial
number of the oracle unitary operations and the nonselective unitary
operations on the initial state (C7), the realizable size of the present NMR
device could be enlarged.

\end{document}